\documentclass{sig-alternate-2013}

\usepackage{amssymb,amsfonts,amsmath}
\usepackage{booktabs}
\usepackage[subrefformat=parens,labelformat=parens]{subfig}
\usepackage{appendix}
\usepackage{balance}
\usepackage{color}
\usepackage{enumitem}
\usepackage[english]{babel}
\usepackage{graphicx}
\usepackage{balance}
\usepackage{xspace}
\usepackage{color} 
\usepackage{caption}
\usepackage{url}
\usepackage{lipsum}

\newcommand{\lyxdot}{.}
\renewcommand{\P}{\mathbb{P}}

\newcommand{\LM}{\text{LM}}
\newcommand{\cC}{\mathcal{C}}

\newcommand{\eat}[1]{}

\begin{document}

\thispagestyle{empty}
\pagenumbering{arabic} 

\numberofauthors{1}

\author{
\alignauthor Bruno Ribeiro\\
       \affaddr{School of Computer Science}\\
       \affaddr{Carnegie Mellon University}\\
       \affaddr{Pittsburgh, PA 15213}\\
       \email{ribeiro@cs.cmu.edu}
}

\title{Modeling and Predicting the Growth and Death of Membership-based Websites}


\maketitle

\begin{abstract}
Driven by outstanding success stories of Internet startups such as Facebook and The Huffington Post, recent studies have thoroughly  described their growth. These highly visible online success stories, however, overshadow an untold number of similar ventures that fail. The study of website popularity is ultimately incomplete without general mechanisms that can describe both successes and failures. In this work we present six years of the daily number of users (DAU) of twenty-two membership-based websites -- encompassing online social networks, grassroots movements, online forums, and membership-only Internet stores -- well balanced between successes and failures. We then propose a combination of reaction-diffusion-decay processes whose resulting equations seem not only to describe well the observed DAU time series but also provide means to roughly predict their evolution. This model allows an approximate automatic DAU-based classification of websites into self-sustainable v.s.\ unsustainable and whether the startup growth is mostly driven by marketing \& media campaigns or word-of-mouth adoptions.
\end{abstract}
\category{G.3}{Probability and Statistics}{Time series analysis}
\category{G.1.7}{Ordinary Differential Equations}{Convergence and stability}

\keywords{DAU; online social dynamics; non-equilibrium statistical\\ physics }



\section{Introduction}
Outstanding  success stories of membership-based websites -- such as online social networks and online forums -- captivate the attention of the general public and research community alike,
inspiring a variety of studies dedicated to the description of the mechanisms behind their rapid growth~\cite{ugander2012structural,backstrom2006group,Jin:2001hi,leskovec2008microscopic,Leskovec07,kairam2012life,Kumar:2010ek}.
But these models only describe the growth in membership (adoptions) rather than the number of members that visit the website per day, henceforth denoted {\em daily active users} (DAU).
This omission is surprising since the DAU is undoubtedly a better proxy of a website's social impact and revenue generation potential than its total number of members~\cite{Coll_Boyd2011,MN_Sornette2012}. The DAU is used to describe the financial health of websites in their quarterly reports, e.g., Facebook~\cite{Facebook}.
Websites with large but overwhelmingly inactive member bases are often deemed ``failures''~\cite{Coll_Boyd2011,Torkjazi:2009it} and have low market values~\cite{MN_Sornette2012}.
Ultimately, even the study of the online social phenomenon~\cite{DavidLazer:2009ci} is incomplete without understanding the mechanisms that can describe the DAU evolution of both successful and unsuccessful websites. 

This study fills this gap with a model that not only captures the DAU evolution observed in 
 high-resolution datasets  -- collected by a leading Internet analytics company -- but also predicts their evolution into the future.
Our data consists of up to six years of daily measurements of the DAUs from twenty-two membership-based websites -- encompassing online social networks, grassroots movements, online forums, political news organizations, and membership-only Internet stores -- totaling over $55,\!000$ website-day observations in a sample well balanced between successful and unsuccessful websites. 
These websites are diverse and in a constant state of change.
Our primary goal is to find a common set underlying dynamics that are immutable throughout the website's life and can roughly describe -- without getting into the particularities of each website -- their membership dynamics, more specifically their DAU time series.

To achieve our goal we present a simple model that uses a set of reaction-diffusion-decay equations describing the attention-seeking interactions between active members, inactive members, and not-yet-members of the website.
The low barrier imposed to subscribe to a  website (often a free service) means that an individual may subscribe to multiple websites~\cite{Goga:2013vv}, showcasing the importance of capturing members attention to the website's survival.
And indeed, we note that today's successful membership-based websites seem to constantly bombard us with attention-seeking messages in the hope that we remain active members, which, as this study shows, is a key factor for their long-term survival.
The end result is a model that embodies Simon's 1969 visionary remarks that, in an information-rich world our attention is bound to become one of our most scarce, important, and vied-for resources~\cite{simon-attention}.

This work is organized as follows.
Sec.~\ref{sec:rel} presents the related work.
Sec.~\ref{sec:mod} presents our proposed model and the algorithm to fit its parameters to the datasets.
Sec.~\ref{sec:res} fits the parameters to the datasets and present the fit and prediction results.
Finally, Sec.~\ref{sec:con} presents our conclusions.

\section{Related work}\label{sec:rel}
Adoption models describe phenomena as diverse individuals deciding to adopt a new technology~\cite{katz1985network,liebowitz1994network,farrell2007coordination,bass2004comments,geroski2000models,fisher1972simple} or individuals adopting a new health habit~\cite{Christakis:2007fy,centola2010spread}.
These models have deeply influenced the study of online social network growth~\cite{ugander2012structural,centola2010spread,backstrom2006group,Jin:2001hi,leskovec2008microscopic,Leskovec07,kairam2012life}.
Roughly, adoption models can be classified as:\\

\noindent
{\bf (a) network effect adoption models} (a.k.a.\ network externality models)~\cite{Garcia:2013ty,katz1985network,liebowitz1994network,farrell2007coordination,Marsili:2004kf,Snijders:1996ea,Jackson:2005tm,Bala:2000do,Skyrms:2009gu,Young:2011br,Montanari:2010cw}, where individual rationality and adoption costs and utilities are modeled in a game-theoretic framework.
Network effect models often assume strategic adopters, i.e., the decision to adopt a product depends on whether the product will gather more adopters in the future. 
Adopters also face adoption and switching costs in a game-theoretic framework~\cite{farrell2007coordination}.
Garcia et al.~\cite{Garcia:2013ty} uses structural features ($k$-core decomposition) and classical network effect theory to predict whether an user departure (node removal) from Friendster triggers other users departures.
Garcia et al.~does not consider the role of user activity (only network structure) and their analysis of Friendster's popularity rests on Google search volume data for searches of ``www.friendster.com'', which is likely an unreliable proxy of the true DAU values.\\

\noindent
{\bf (b) threshold adoption models}~\cite{granovetter1978threshold,schelling2006micromotives}, where an individual adopts if enough of his or her friends are adopters;
The threshold models~\cite{granovetter1978threshold,schelling2006micromotives} describe product (and behavior) adoptions in social networks, where an individual adopts a product if ``enough'' of his or her friends are adopters.
Adopters are not strategic and have only a local view of the network.
The notion of how many friends are ``enough'' vary from individual to individual and may also be determined by the number of friends and their interconnections~\cite{ugander2012structural,centola2010spread}.\\

\noindent
{\bf (c) diffusion of innovation models}~\cite{bass2004comments,geroski2000models,fisher1972simple}, where adopters influence others to adopt through word-of-mouth; In the absence of fine-grained individual-level data these models provide demand forecasting at the aggregate (population) level.
The Bass~\cite{bass2004comments,geroski2000models} and the Fisher-Pry~\cite{fisher1972simple} models are arguably two of the best known models of diffusion of innovations.
In the Bass model, the diffusion of a product is driven by two forces: word-of-mouth (where adopters recruit new adopters) and marketing (marketing campaigns recruit new adopters).
Diffusion of innovation models are also known as logistic models because of the S-shaped curve of the number of adopters over time, which replicates the adoption curve of a variety of real world measurements~\cite{Mansfield61,bass2004comments,Mansfield63}.
Diffusion models have also been applied in the field of sociology to describe diffusions of social behavior (see Strang and Soule~\cite{strang1998diffusion} for a review).\\

\noindent
{\bf (d) adoption models from influence and network structure}~\cite{ugander2012structural,centola2010spread,backstrom2006group,Jin:2001hi,leskovec2008microscopic,Leskovec07,kairam2012life,Kumar:2010ek}, where an individual adoptions depends not only on whether his or her friends adopt but also on how these friends are connected among themselves.
Threshold models have inspired a variety of empirical research~\cite{ugander2012structural,centola2010spread,backstrom2006group,leskovec2008microscopic}.
The probability that a non-member user joins the network is known to increase linearly with the number of invitations if the number of invitations is small, suffering a diminishing return effect as the number of invitations grow~\cite{ugander2012structural,centola2010spread,backstrom2006group}.
These findings would have a measurable effect in our problem formulation if most susceptible users were to receive a large number of invitations prior to joining the website, which is often not quantified~\cite{ugander2012structural,backstrom2006group} or is a parameter of the experiment~\cite{centola2010spread}.
A variety of works also consider the relationship between community growth inside an online social network (OSN) websites and their network structure~\cite{kairam2012life,backstrom2006group,leskovec2008microscopic}.
These studies, however, focus on (i) the growth of communities inside the OSN (not the growth of the OSN itself) and (ii) the role of network structure disregarding whether the community is alive (active) or dead (inactive).

Other works (that do not quite fit in the above categories) consider the popularity of entities such as news items and videos (interestingly, these items are measured at widely successful websites)~\cite{Wu:2007ej,Dezso:2006eq,Szabo:2010jl,Crane:2008hm}.
These works measure the total number of news readers or video viewers and, thus, can also be considered in the realm of adoption models because news readers generally do not reread the same article or re-watch the same video.
In contrast, the mechanisms that drive members to return to a website depend on the complex interactions between non-members, new members, and active and inactive members.

\paragraph{Modeling DAU dynamics}
The above adoption models only describe the evolution of the total number of members.
Modeling the DAU evolution requires the modeling of the attention-seeking interaction between active members and other Internet users.
Recent Facebook shows that the activity of our Facebook friends in the website incites us to login and become active which, in turn, incites our friends to either become active or stay active~\cite{backstrom2011center}.
This corroborates with our understanding that active members incite inactive members into activity, which should be one of the key driving forces behind a model of DAU evolution.
We, however, are interested in membership-based website in a broader sense than just online social networks.
Be it a political news website, an online social network, or a grassroots movement, the interactions between active members, inactive members, and non-members is key to the understanding of the DAU dynamics.

The work that most closely resembles ours is Cauwels and Sornette~\cite{MN_Sornette2012} which focuses on describing the evolution of the Facebook DAU.
Cauwels and Sornette, however, is incomplete in the sense that its time-series analysis is tailored towards successful websites and cannot capture sudden DAU drops observed in unsuccessful websites.
The question of what takes to increase the DAU is of great interest to the industry.
Many complex factors help determine when and why members join and leave a website such as cultural and racial trends~\cite{Coll_Boyd2011}.
In this work, however, we do not analyze or model complex societal interactions and trends.
Rather, we opt for a population-level model which we believe can be used as a foundation to build upon, providing valuable insights into the relationship between member activity, inactivity, and website growth, as seen next.

\section{Proposed Model} \label{sec:mod}

Models of interacting populations have been successfully applied in fields as diverse as mathematical biology~\cite{murray2002mathematical}, economics~\cite{Friedman:1991wr}, operations research~\cite{Mansfield61,Mansfield63}, and marketing~\cite{bass2004comments,geroski2000models,rogers1995diffusion}.
Our model considers a large user population and the absence of sudden unpredictable events -- such as a new round of VC funding that allows the website to significantly improve its appeal or increase its media \& marketing exposure, or even the presence of strong new  competitors -- we model the population interactions as coupled reaction, decay, and diffusion processes.
Reaction, diffusion, and decay processes find applications in chemistry, physics, and applied mathematics~\cite{Colizza:2007ik,Colizza:2008fk,vanWijland:1998jl,Kampen:1981vs,murray2002mathematical}.
{\em We choose to avoid stochastic models (which would allow us to give confidence intervals to our predictions) 
because very little is known about the stochastic behavior of the dynamics between inactive and active members of websites and between the latter and non-members}.

Our model can be described as follows.
Consider the interactions between an active member $A$, an inactive member $I$, and a non-member $U$ as reactions, decays, and diffusions in a solution.
The first {\bf reaction}
\[
A+I \to 2A, \label{e:AI2A}
\]
describes the reaction with constant rate $\alpha$ as the marginal influence of active members to attract an inactive members to become active.
This constant reaction approximates two real world phenomena: (a) the activity of website members, such as posting pictures, events, messages, sending links to their friends, writing about their activity on online forums, instant \& e-mail messages, face-to-face interactions, etc.,  attract the attention of inactive members making them active again\footnote{\small
The reaction effect does not take into consideration the impact of communication constraints between members. This effect can be easily incorporated, see our older technical report~\cite{techreport} for a model;
Additionally, over half of the websites analyzed in this study provide no explicit connectivity among their members, 
reducing the utility of reactions tailored to website-specific communication constraints in our analysis.
} (as empirically observed on Facebook~\cite{backstrom2011center}); and (b) the marginal increase in website utility as it gains more active members, an effect known as {\em network effect} or {\em network externality} discussed at length in Farrell and Klemperer~\cite{farrell2007coordination}. 
There are many types of network effect, but the most widely used effect in its purest form can be described the following {\em path-dependent cumulative return} rule (see Arthur~\cite{Arthur} for more details): ``more buyers'' $\to$ ``lower prices'' $\to$ ``more buyers'' or, alternatively, in a reaction better suited to our scenarios: higher DAU $\to$ more advertisement revenue $\to$ better website features $\to$ less inactive members (increased DAU).

Members do not remain active indefinitely. 
Active members eventually become ``temporarily'' inactive. The {\bf decay}
\[
A \to I, \label{e:AI}
\]
describes an active member spontaneously becoming inactive with constant  decay rate $\beta$.
The second {\bf reaction }
\[
A+U \to 2A
\]
happens when an active member influences a non-member to join the website with constant reaction rate $\gamma$, which can happen either through word-of-mouth or because 
of increased utility (e.g., {\em network effects}),  two widely known phenomena in the specialized literature~\cite{bass2004comments,rogers1995diffusion,farrell2007coordination}.
Finally, the {\bf diffusion}
\[
U \to A
\]
describes the diffusion of marketing and media campaigns over non-members with diffusion rate $\lambda$, which then influences the non-member to join the website.

Another parameter of our model is $C$, the fraction of the active Internet population that is targeted by the website service.
In mathematical biology $C$ is known as the carrying capacity.
In our model we partition $C$ into active members, $A(t)$, inactive members, $I(t)$, and non-members, $U(t)$, which are part of the target niche that have not yet joined the website at time $t$ such that $A(t) + I(t) + U(t) = C$.
As $C$ is fraction of the active Internet population we do not need to add extra parameters to account for Internet population growth or the natural seasonal variability in the number of active members due to national holidays or school holidays.
Gathering all these effects and collecting all the terms yields the following set of equations
\begin{equation} \label{eq:SAI}
\begin{aligned}
\frac{dA(t)}{dt} & = -  \frac{1}{C} A^2(t)\gamma  + \frac{1}{C} I(t)A(t)(\alpha - \gamma) + C \lambda \\
&\quad -A(t)(\beta + \lambda - \gamma) -I(t) \lambda    \, , \\ 
\frac{dI(t)}{dt} & = A(t)\beta  -  \frac{1}{C}I(t)A(t)\alpha  \, ,
\end{aligned}
\end{equation}
that describes the dynamics of $A(t)$ which is the DAU as a fraction of the active Internet population (``\%DAU'') at time $t$.
In the {Appendix} we see that the website joining rates and the size of the inactive population prescribed by the above model are consistent with metrics observed at large membership-based website.
~\\

\noindent
{\bf N.B.\ I:} An earlier version of our technical report~\cite{techreport} provides model extensions that account for multiple marketing and media campaigns, growth in the target population $C$,  account cancellations, and  the sudden appearance of predatory competition by other website (e.g. the competition of Facebook against MySpace, Orkut, Hi5, Friendster, and Multiply).
These models are left out of this paper due to space constraints.

\subsection{DAU long-term activity} \label{sec:lt}

After a certain -- possibly large -- time $t^\star$ the website nearly exhausts its pool of non-members and growth happens only when the Internet user population grows, i.e., $U(t) \approx 0$ for all $t \geq t^\star$.
After time $t^\star$ the DAU must be sustained through the Active $\leftrightarrow$ Inactive dynamics.
For $t > t^\star$ we can then approximate $I(t) \approx C - A(t)$; substituting in~\eqref{eq:SAI} we obtain
\begin{align*}
\frac{dA(t)}{dt}& \approx -A(t)\beta+(1 - A(t)/C)A(t)\alpha. \label{eq:eql}
\end{align*}
and, thus, for $t > t^\prime$,
\begin{equation} \label{eq:Asus}
A(t) \approx \frac{(\beta - \alpha) e^{-(\beta - \alpha)t^\prime}}{1 -  \frac{\alpha}{C} e^{-(\beta - \alpha)t^\prime}},
\end{equation}
where $t^\prime = t +h$, with $h$ a constant.
In equilibrium
$$
\lim_{t \to \infty} A(t) = \begin{cases}
                     0 & \text{if } \beta/\alpha \ge 1,\\
                     C(1 - \beta/\alpha) & \text{if } \beta/\alpha < 1.
                     \end{cases}
$$
Equation~\eqref{eq:Asus} predicts that any website whose ratio $\beta/\alpha \ge 1$ will eventually decay into inactivity, regardless of how much marketing and media exposure the website gets or how strong the word-of-mouth buzz is.
More importantly, we can predict the function that describes its decay.
The website fades into inactivity as $A(t) \propto e^{-(\beta - \alpha)t}$.
In what follows we see that these predictions match surprisingly well the observed data.

Note that an increase in $\alpha$ has a great impact on the number of active members.
Not surprisingly, online social network websites have recently been targeting member inactivity. 
For instance, Facebook in recent years has introduced notification messages of the form: ``Here's some activity you may have missed'' (Facebook members may opt out of such messages), even if almost no member activities are reported or these emails are consistently ignored by the user.
Network growth is not the only reason why online social networks want to keep their members active. 
The website revenue is often tied to its DAU~\cite{MN_Sornette2012}, mostly through online ads.
But, most importantly, our model shows that the website survival depends on whether the ratio $\beta/\alpha \geq 1$, creating a great incentive to increase $\alpha$ through ``here's some activity you may have missed'' reminders.

\begin{figure}[!!ht]
\centering
\includegraphics[width=3.4in,height=1.3in]{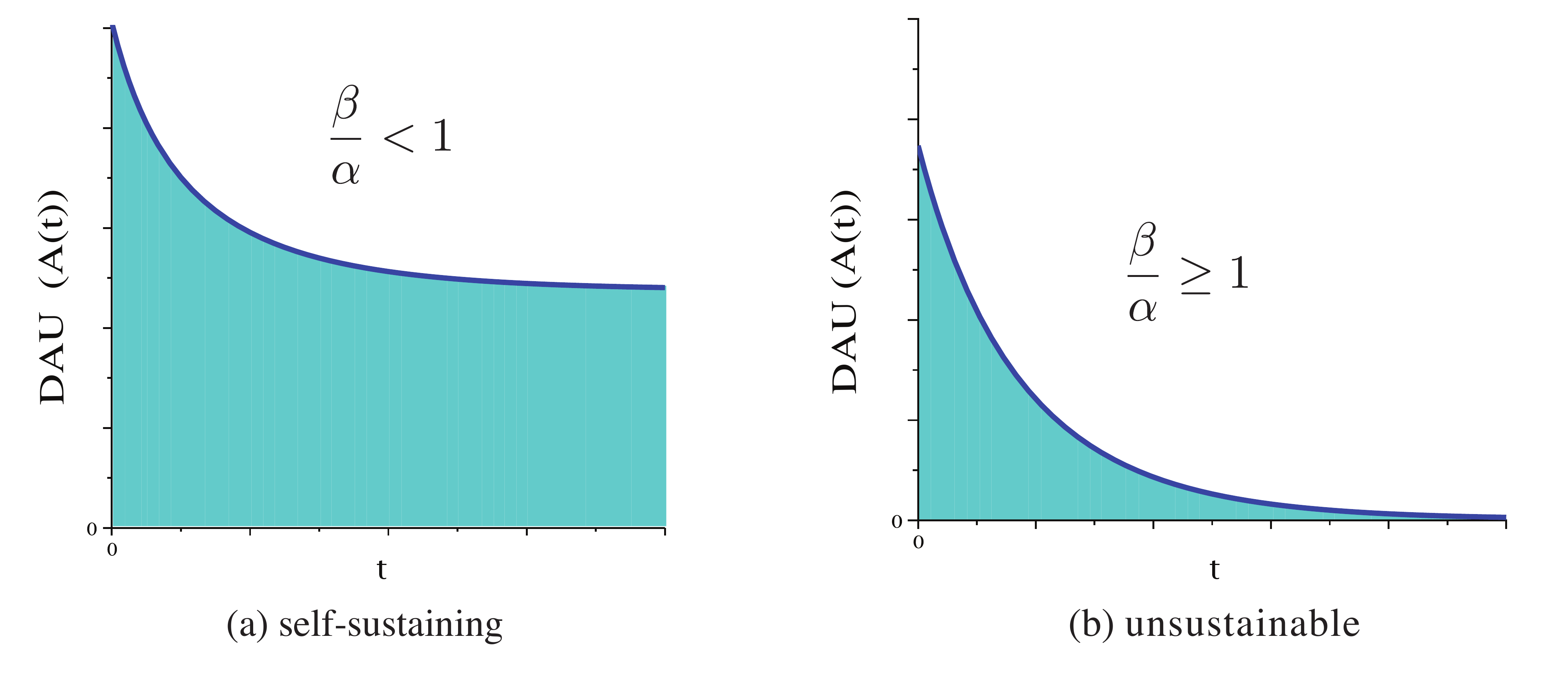}
\caption{Long-term DAU signatures\label{f:asig}}
\end{figure}

{\bf The DAU signature of equation~\eqref{eq:Asus}.} 
Equation~\eqref{eq:Asus} predicts two distinct DAU signatures related to the coupling of activity $\leftrightarrow$ inactivity reactions and activity decay.
Fig.~\ref{f:asig} shows two likely DAU signatures.
The {\bf self-sustaining} DAU behavior in Fig.~\ref{f:asig}(a) is observed when $\beta/\alpha < 1$ and the initial DAU is higher than the asymptotic DAU level of $C(1 - \beta/\alpha)$. The curve shows a slow decay from the initial DAU towards $C(1 - \beta/\alpha)$.
The {\bf unsustainable} (decaying) DAU time series in Fig.~\ref{f:asig}(b) is observed when $\beta/\alpha \geq 1$ and always converges to zero irrespective of the starting DAU value.

\noindent
{\bf N.B.\ II:} The  set of equations in~\eqref{eq:SAI} seems to be at the right level of complexity.
For instance, consider an alternative simpler model where the reaction term $I(t)A(t)\alpha/C$ in~\eqref{eq:SAI} is replaced by a simpler diffusion term $I(t)\zeta$.
That is, inactive members become active due to external factors and not as a function of the current active member base as in~\eqref{eq:SAI}.
The simpler model predicts that member activity always asymptotically reaches $C/(1 + \beta/\zeta)$, never fading.
But such prediction simultaneously defies common sense and the available data.
In the Appendix we see an example where the simpler model is unable to match the DAU time series observed in our datasets.
Increasing the complexity of~\eqref{eq:SAI} by having both $I(t)A(t)\alpha/C$ and $I(t)\zeta$ terms also has a detrimental effect. 
The addition of the extra parameter $\zeta$ leads to better model fit but also to less accurate predictions that do not have the DAU fade to zero when $\beta/\alpha \geq 1$.

\subsection{DAU signatures of growth}\label{sec:gr}
Intense word-of-mouth and media \& marketing effects leave distinctive signatures.
From a small number of active members an intense word-of-mouth suffers exponential growth in the initial DAUs ramp up phase due to the word-of-mouth reaction process feedback higher DAU $\to$ word-of-mouth growth $\to$ higher DAU.
On the other hand, media \& marketing diffusions have a characteristic convex growth that is independent of the current DAU value.

\begin{figure}[!!ht]
\centering
\includegraphics[width=3in,height=1.3in]{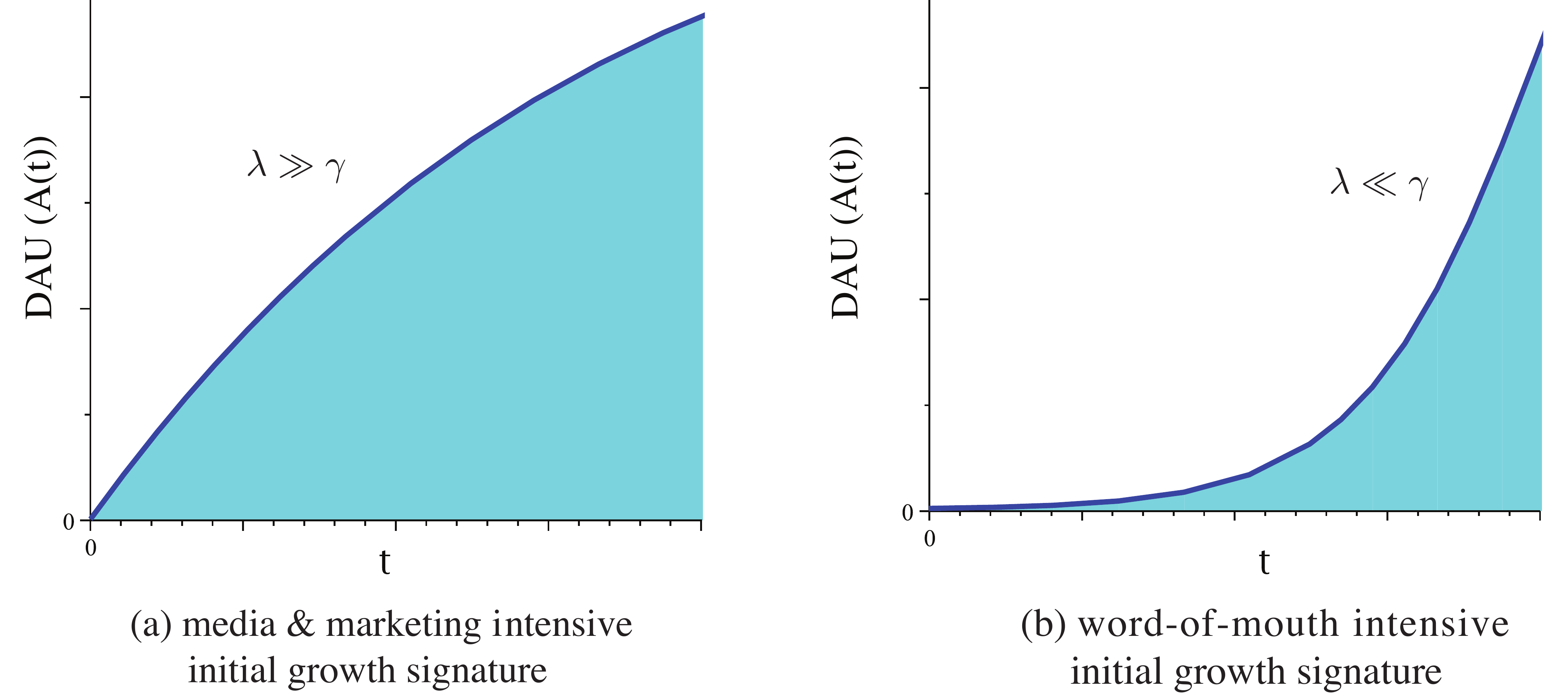}
\caption{DAU signatures of initial membership growth.\label{f:ado}}
\end{figure}

In  websites that depend heavily on media \& marketing campaign growth has $\lambda \gg \gamma$ while in websites 
that depend heavily on word-of-mouth growth has $\lambda \ll \gamma$.
For $t^\star$ small such that $\forall t < t^\star$ we have  $U(t) \approx 1$.
When $\lambda \gg \gamma$ (media \& marketing intensive adoptions) equation~\eqref{eq:SAI} yields
\[
\frac{dA(t)}{dt} \approx   (C -A(t)) \lambda, \quad \forall t < t^\star, \: \lambda \gg \gamma.
\]
Fig.~\ref{f:ado}(a) shows an initial DAU time series signature of websites with media \& marketing intensive adoptions.
Similarly, when $\lambda \ll \gamma$ (word-of-mouth intensive adoptions) equation~\eqref{eq:SAI} can be approximated
\[
\frac{dA(t)}{dt} \approx  A(t)\gamma \left(1  - \frac{A(t)}{C}\right), \quad \forall t < t^\star, \: \lambda \ll \gamma.
\]
Fig.~\ref{f:ado}(b) shows an initial DAU time series signature of websites with word-of-mouth intensive adoptions.

The analysis of Secs.~\ref{sec:lt} and~\ref{sec:gr} show some isolated effects instead of all coupled interactions in our model.
In what follows we fit the parameters of our model to the DAU time series of twenty-two websites and compare the full dynamics of the DAU time series predicted by equation~\eqref{eq:SAI} and the real DAU time series.

\begin{figure*}[!!t]
\setcounter{subfigure}{0}
\captionsetup{type=figure}
\centering
\hfill
\subfloat[][{\bf Sustainability:} fitted inactive$\to$active reaction rate $\alpha$ v.s.\ inactivity decay rate $\beta$.\label{f:ab}]{
\hspace{-.15in}
\includegraphics[width=3in,height=2.5in]{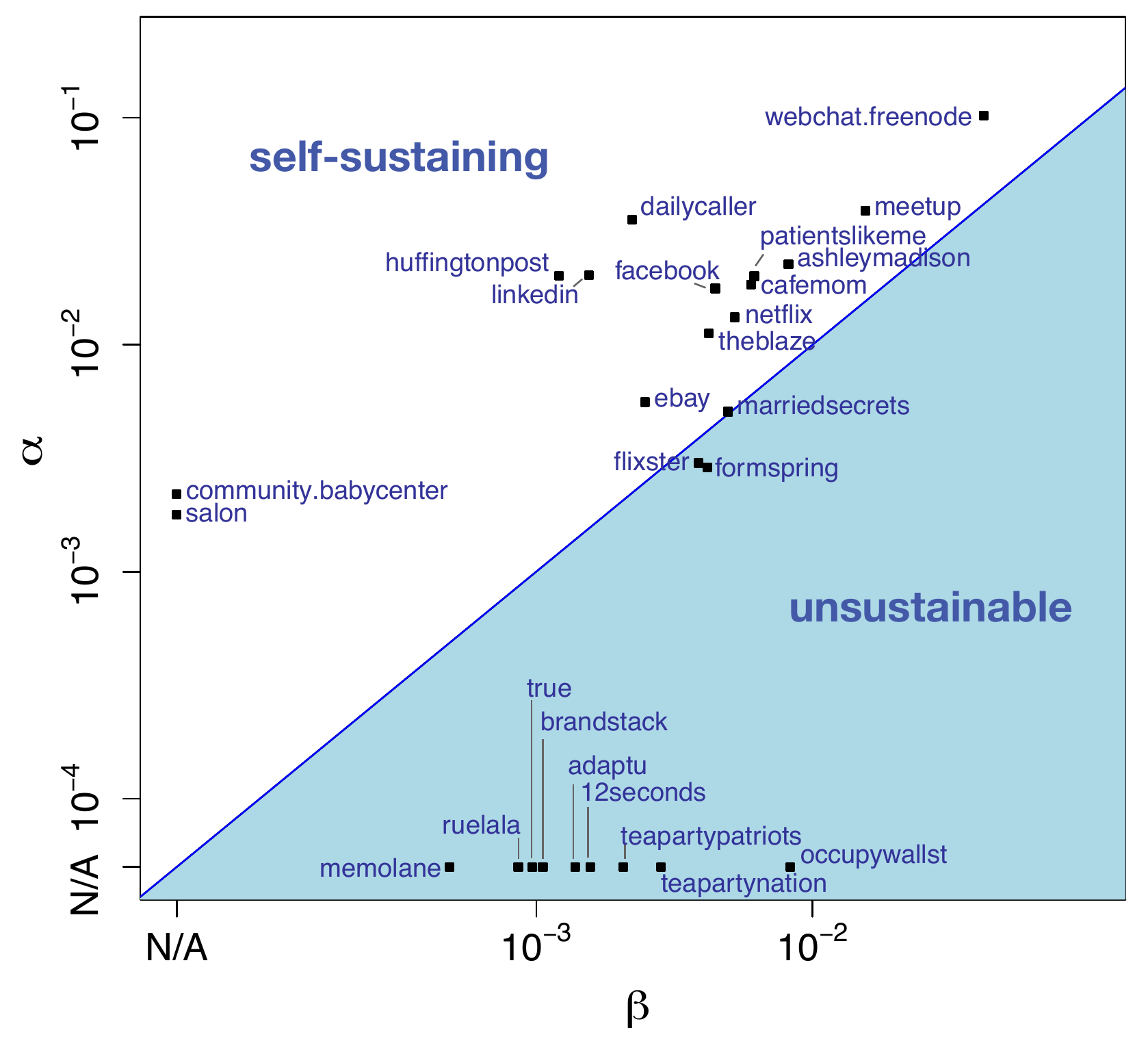}
}
\hfill
\subfloat[][{\bf Growth:} fitted  marketing \& media diffusion rate $\lambda$ v.s.\ word-of-mouth reaction rate $\gamma$.\label{f:mw}]{
\includegraphics[width=3in,height=2.5in]{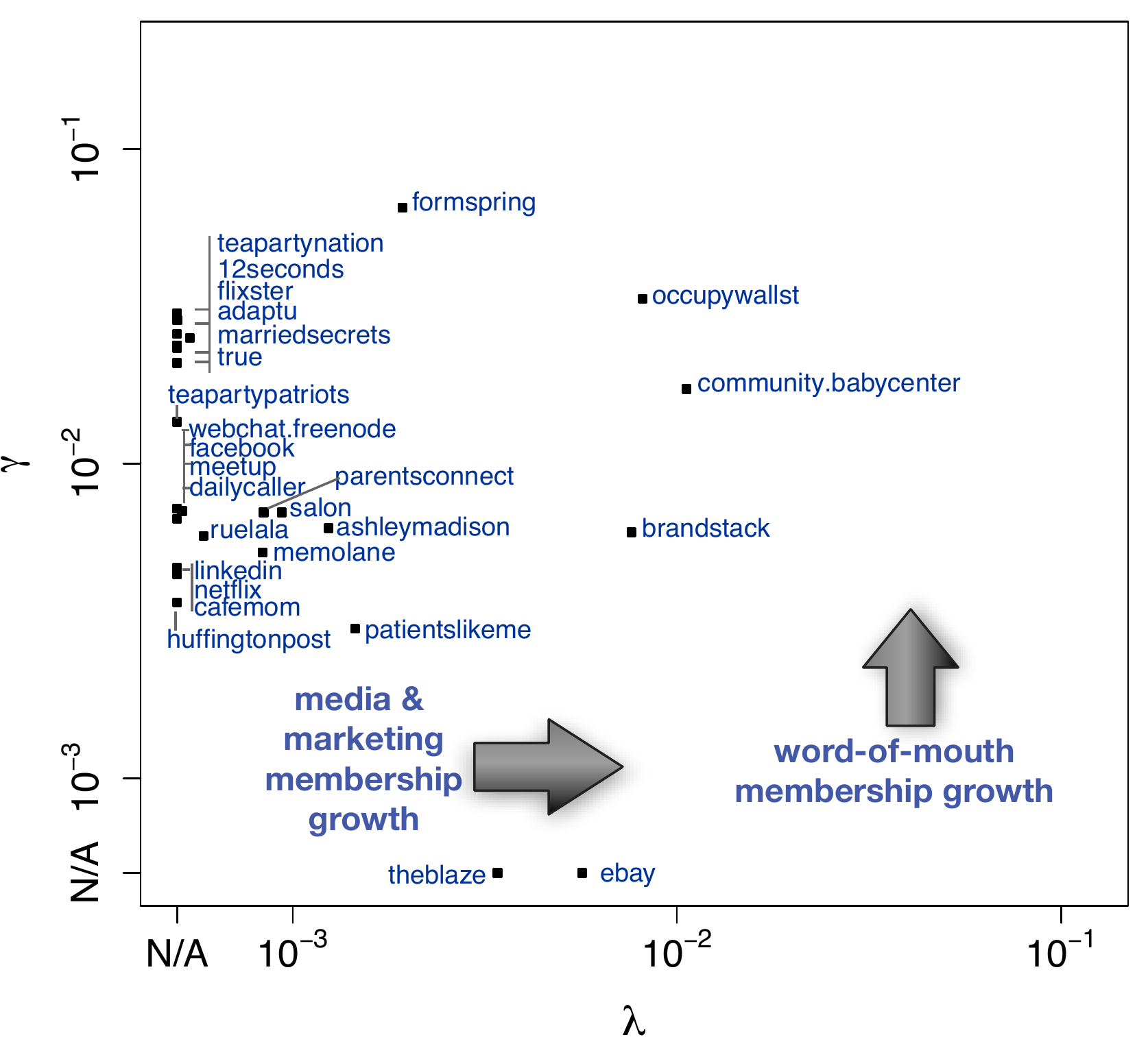}
}
\hfill
\caption{Website classification according to automatically fit parameters.\label{f:lambda_gamma}}
\end{figure*}


\section{Results} \label{sec:res}
Our results section has two objectives, namely:
\begin{enumerate}
\item Verify whether our model can be fit to the DAU time series observed in our datasets and whether the DAU time series behavior predicted by our model is observed in the datasets.
\item Automatically fit the model parameters (without fiddling with initial conditions) to a training set of the data consisting of between $25\%$ and $60\%$ of the initial DAU time series, reserving the rest of the DAU data as a {\em test set} to test the DAU predicted by our fitted model.
\end{enumerate}

\subsection{Datasets}
The DAU data was measured from June 2007 to June 2013 (six years) as a fraction of the active Internet population of twenty-two websites.
The data is provided by Amazon's Alexa web analytics company totaling more than $40,\!000$ website-day observations.
The chosen websites encompass online social networks, online forums, political news websites, membership-based retail stores, and online social movements, all ranging from outstanding successes to catastrophic failures.
We choose websites that either require membership (e.g., Facebook, Netflix, LinkedIn) or depend on the activity of a loyal user base (.e.g, the left-leaning news aggregator TheHuffingtonPost).  
Unsuccessful websites urls were collected from TechCruch's  failed startup's epitaphs~\cite{deadpool} between February and August of 2012 and 2013 while successful websites names were gathered in the press and the related literature.
As standard practice we smooth out the DAU outliers using a moving median with a 31-day DAU interval centered around each day.
A description of the websites and further details about the data collection procedure are found in the Appendix.

It is important to note that Alexa's datasets do not include smartphone traffic.
Therefore we do not consider websites that have significant smartphone-based traffic (e.g., Twitter, GaiaOnline, RockYou).
Facebook is the only exception because our Alexa's DAU metrics seem to have suffered little from the introduction of the Facebook smartphone app, i.e., despite the smartphone app Facebook users seem to still also access the website through their desktop or laptop computers.

\subsection{Automatic parameter fit}
\begin{figure*}[!htb]
\setcounter{subfigure}{0}
\captionsetup{type=figure}
\hspace{.1in}
\centering
\subfloat[][Common unsustainable + word-of-mouth DAU shape.\label{f:tunwm}]{
\includegraphics[width=1.6in,height=1.3in]{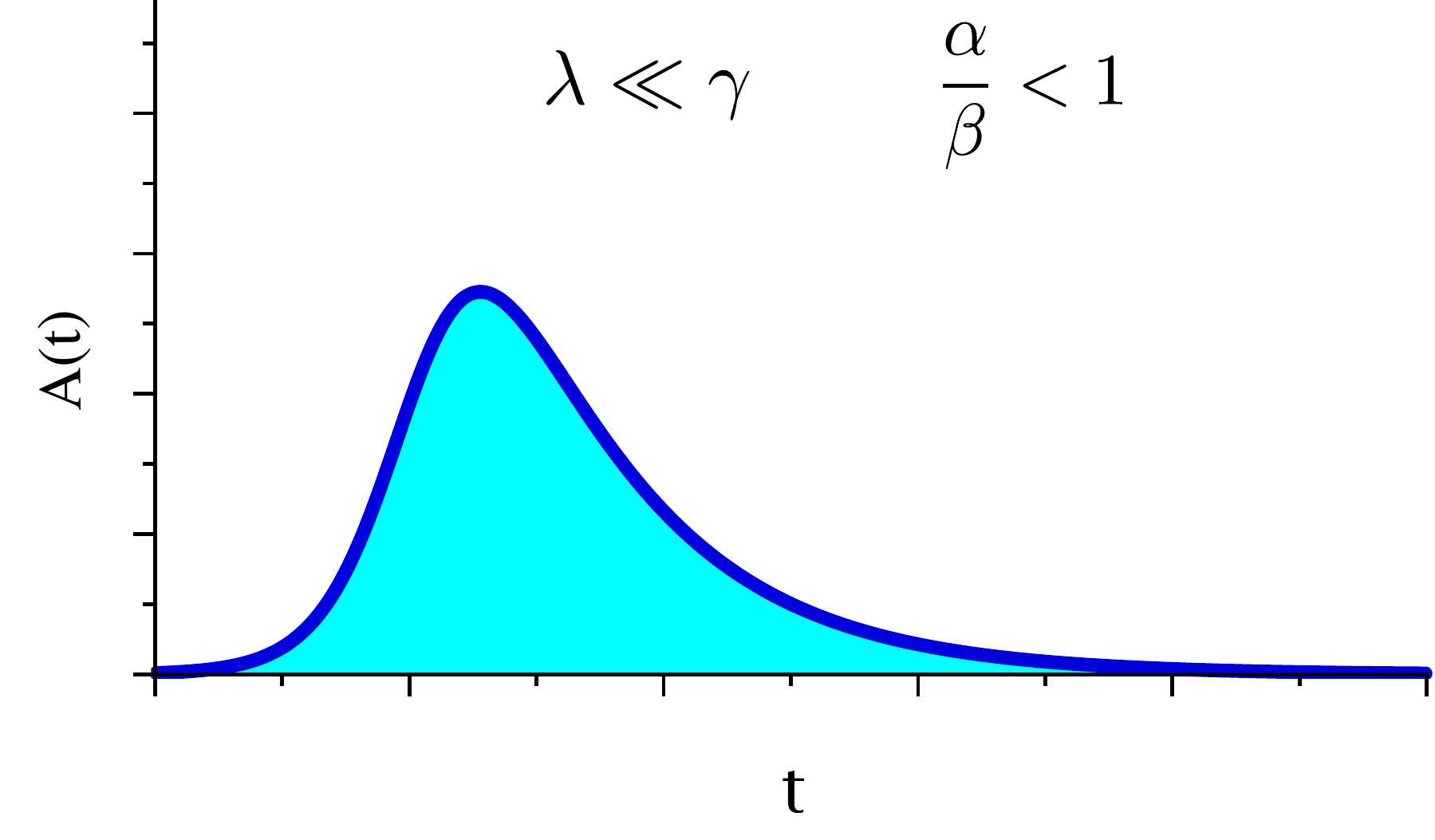}
}
\subfloat[][Common unsustainable + media \& marketing exposure DAU shape.\label{f:tunmk}]{
\includegraphics[width=1.6in,height=1.3in]{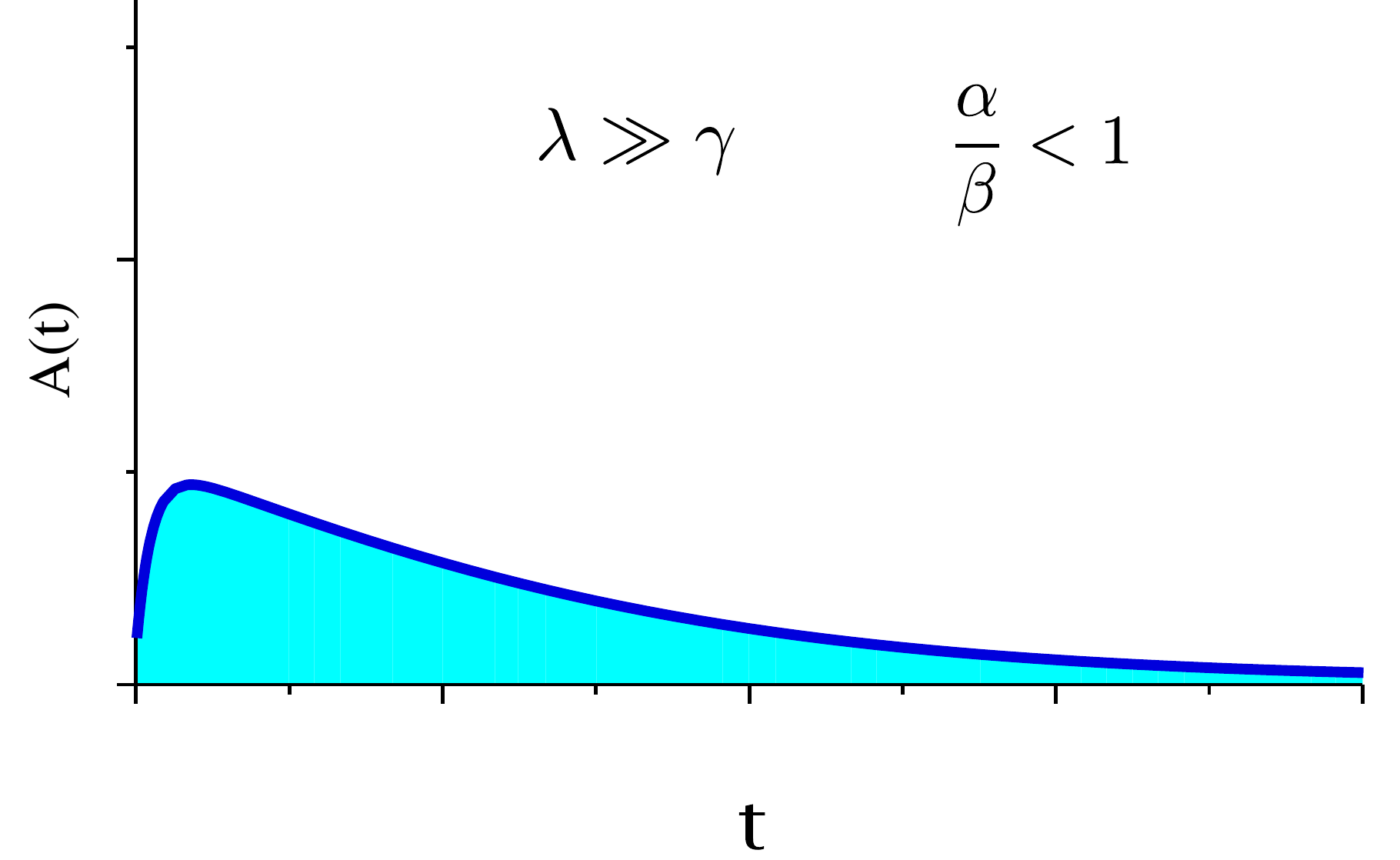}
}
\subfloat[][occupywallst.org\label{f:occ}]{
\includegraphics[width=1.6in,height=1.3in]{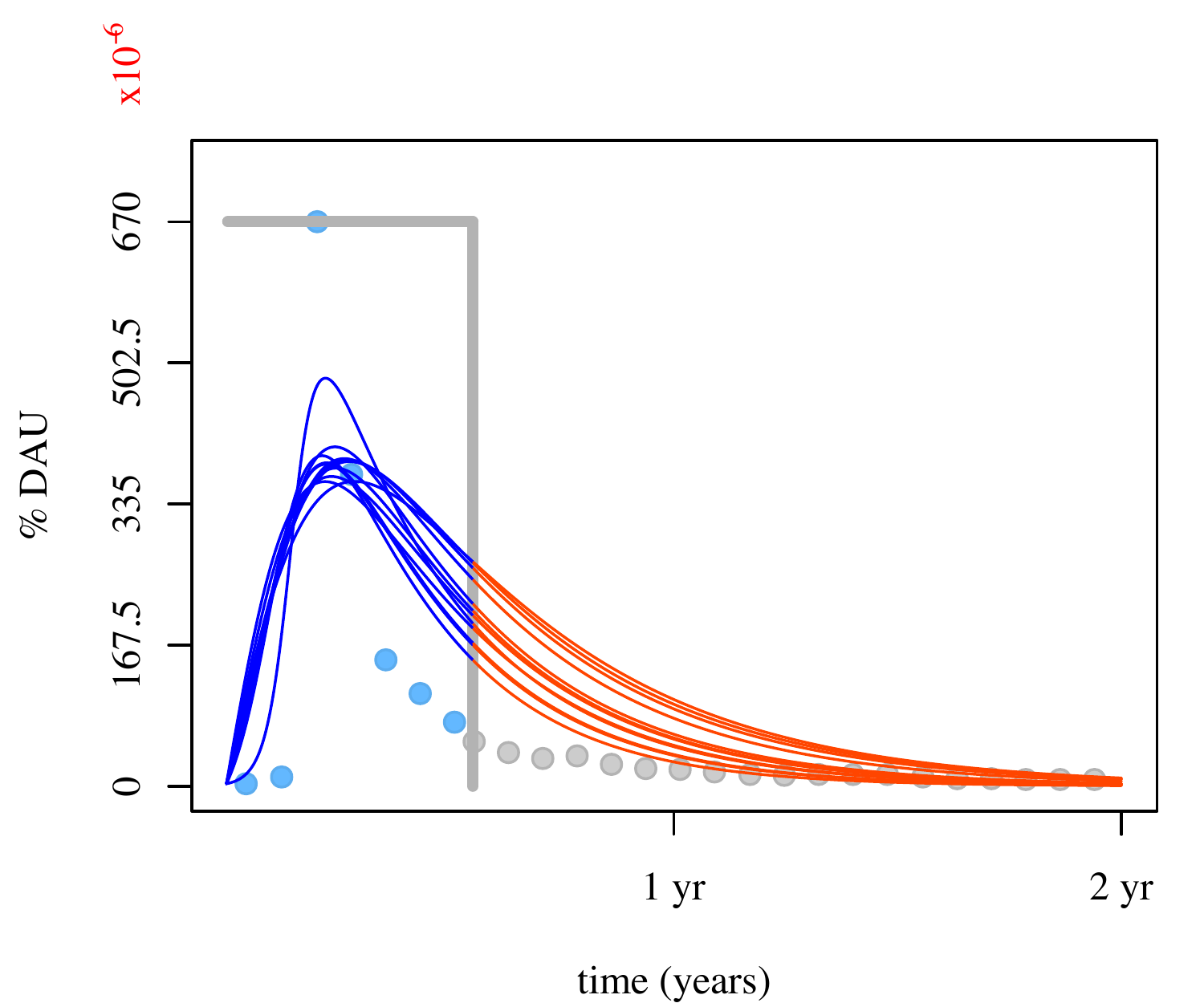}
}
\subfloat[][12seconds.tv\label{f:12s}]{
\includegraphics[width=1.6in,height=1.3in]{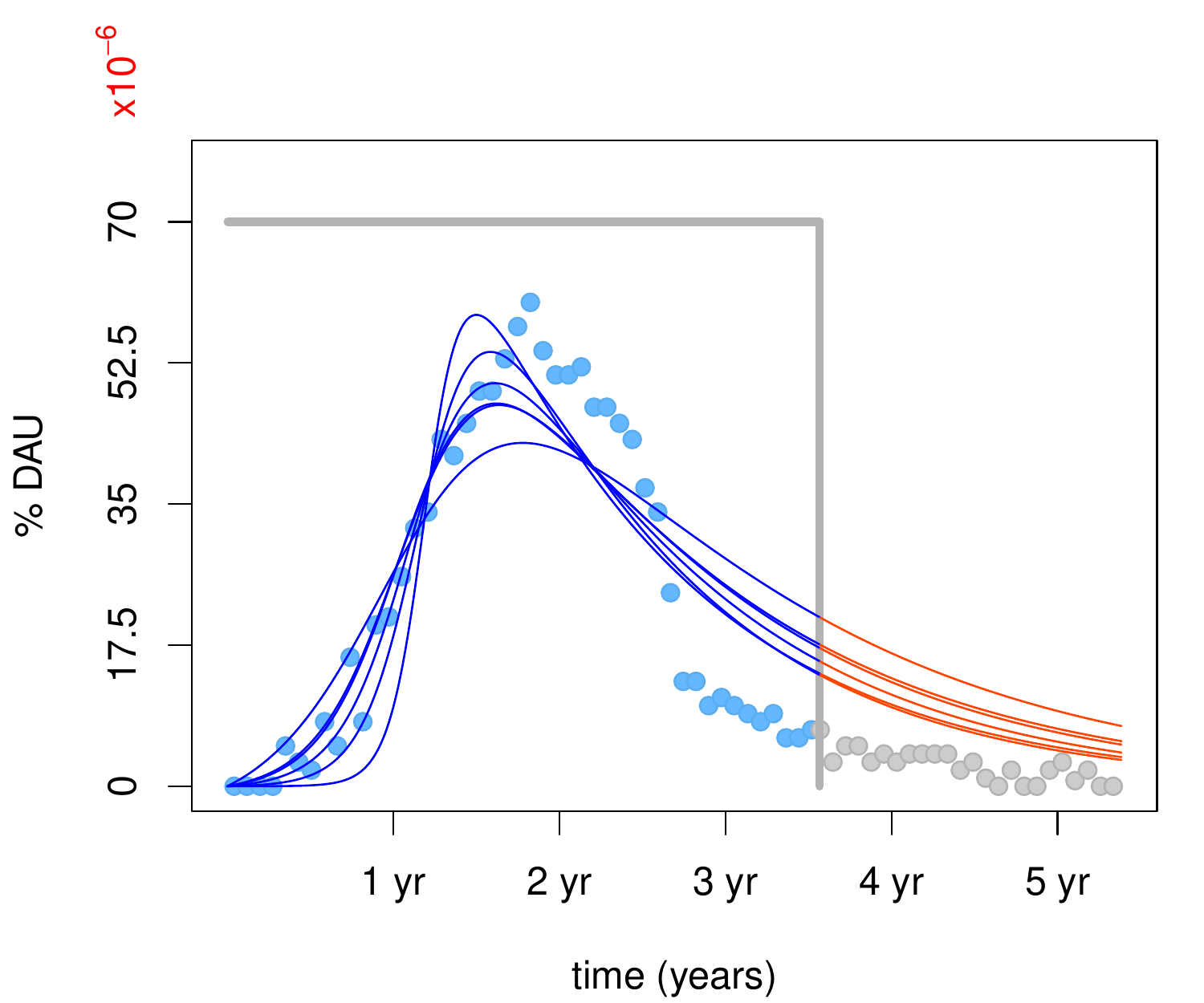}
}
\\
\subfloat[][formspring.me\label{f:for}]{
\includegraphics[width=1.6in,height=1.3in]{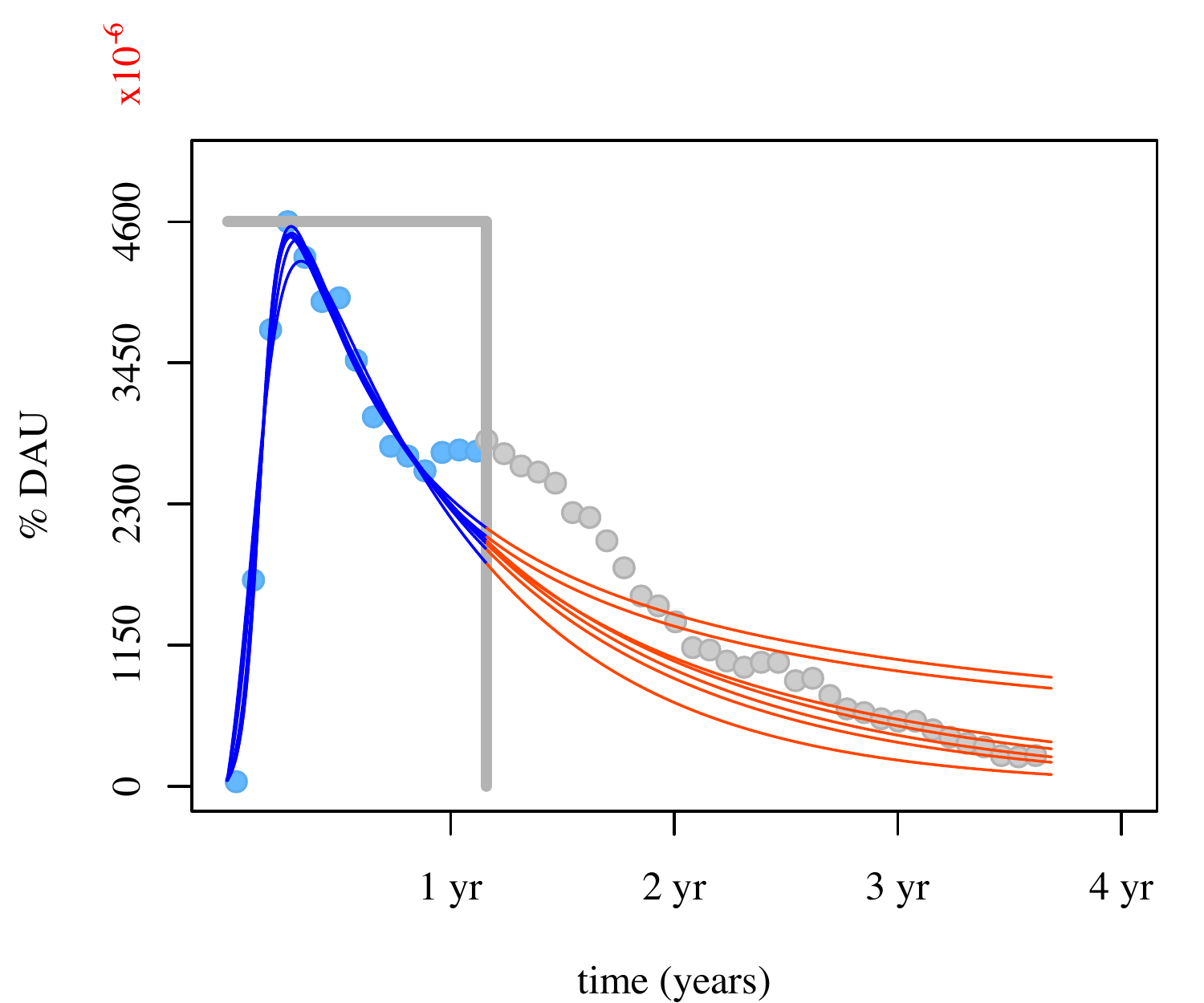}
}
\subfloat[][ruelala.com\label{f:rue}]{
\includegraphics[width=1.6in,height=1.3in]{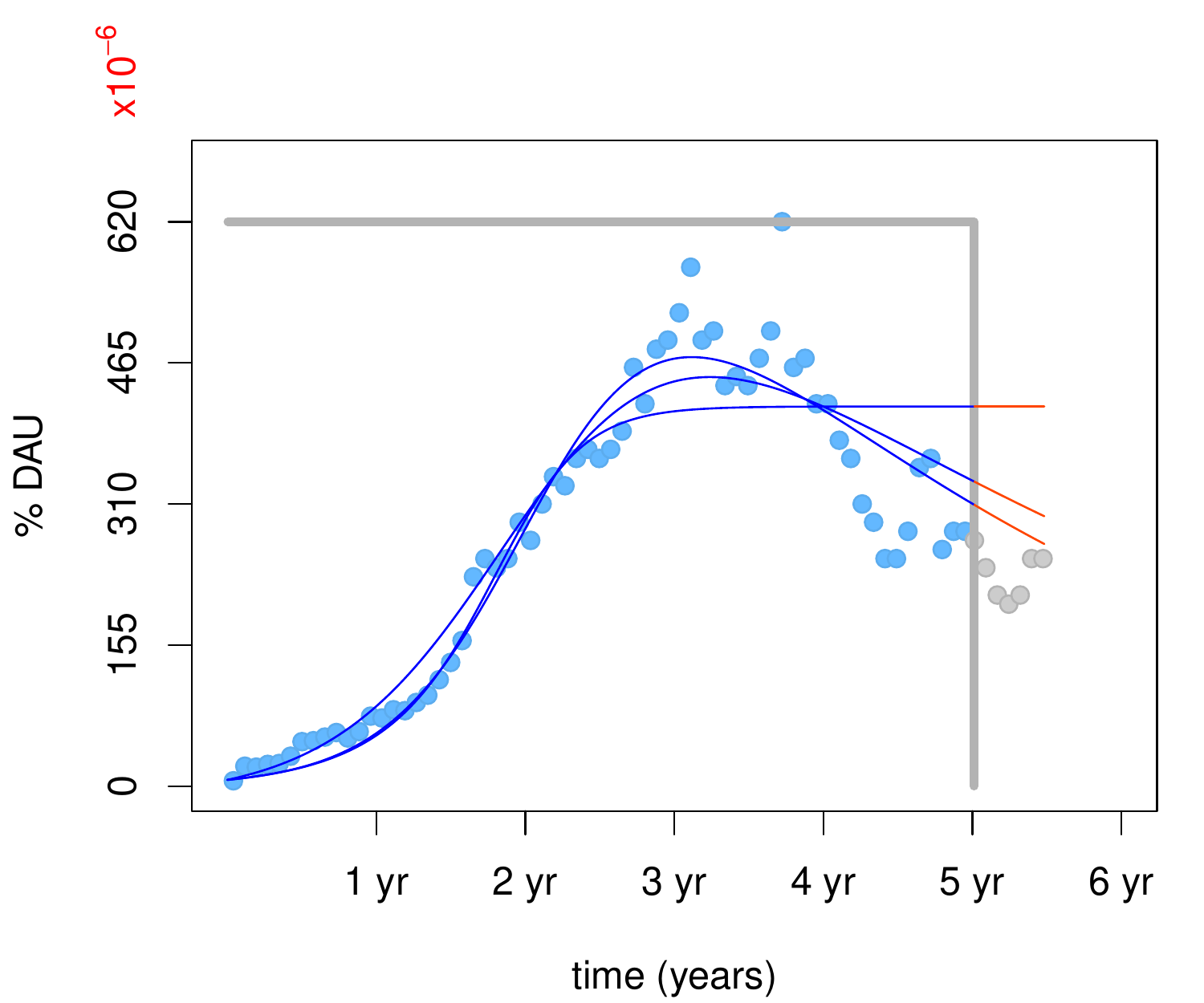}
}
\subfloat[][adaptu.com\label{f:ada}]{
\includegraphics[width=1.6in,height=1.3in]{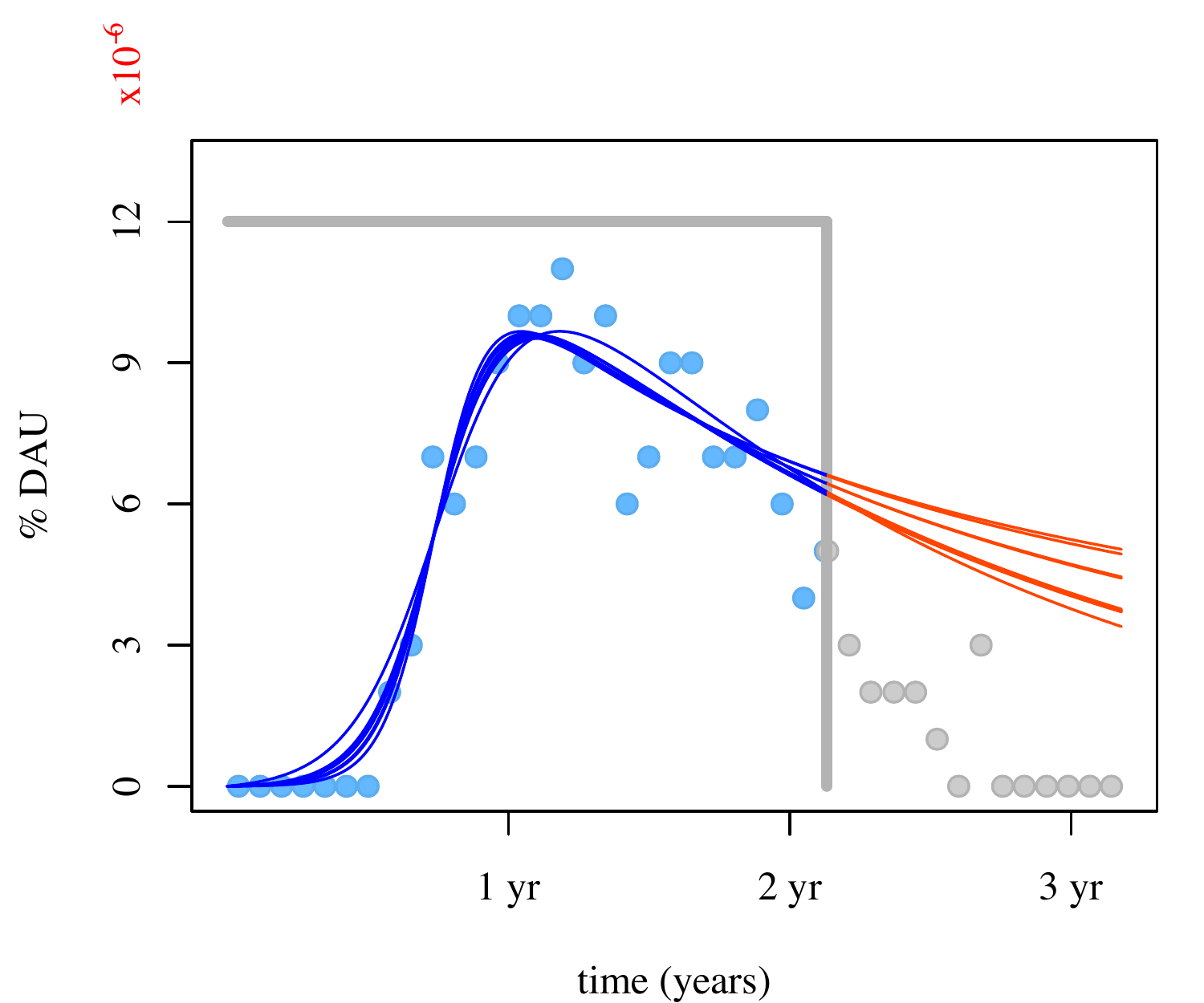}
}
\subfloat[][flixster.com\label{f:fli}]{
\includegraphics[width=1.6in,height=1.3in]{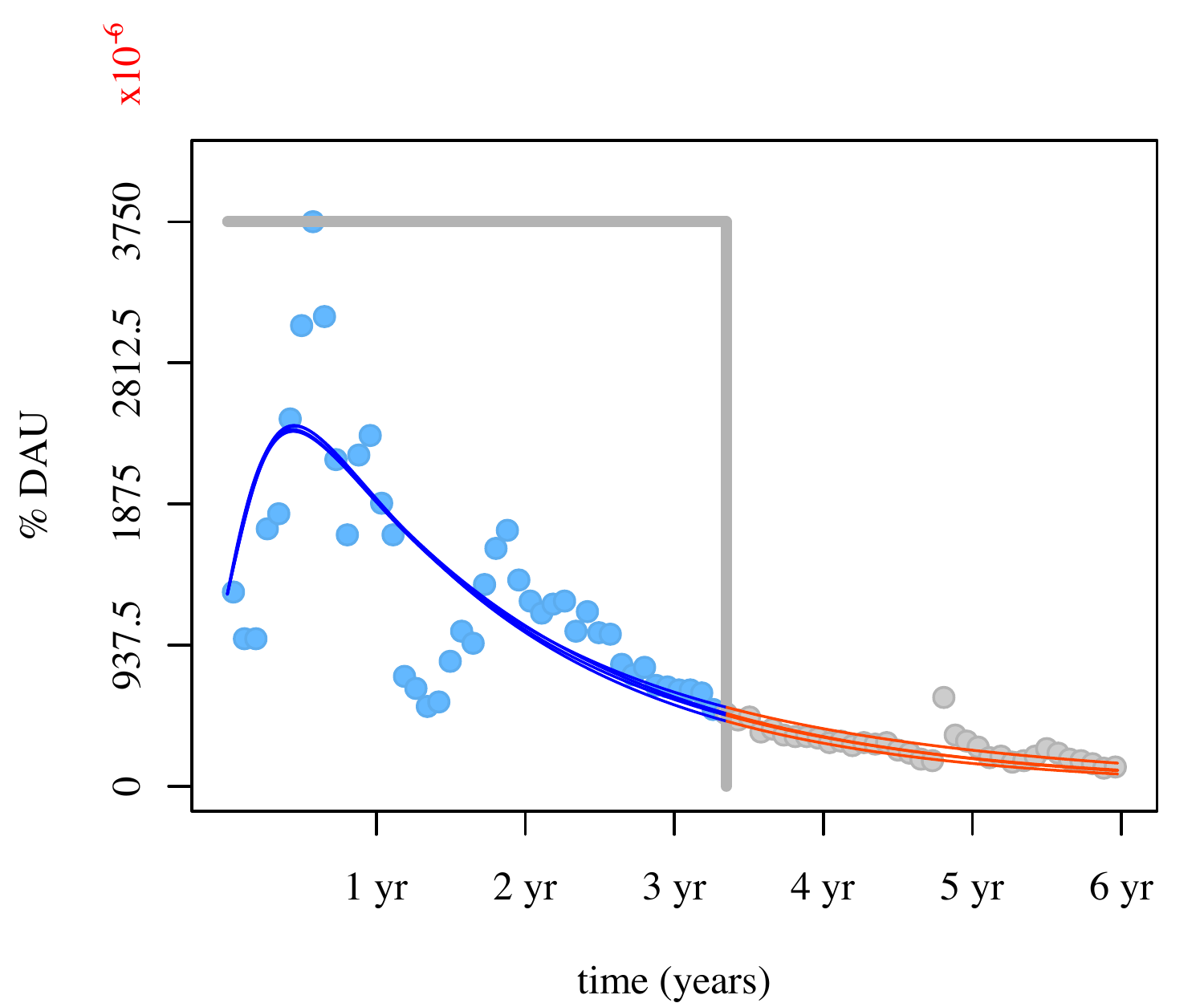}
}
\\
\subfloat[][teapartynation.com\label{f:par}]{
\includegraphics[width=1.6in,height=1.3in]{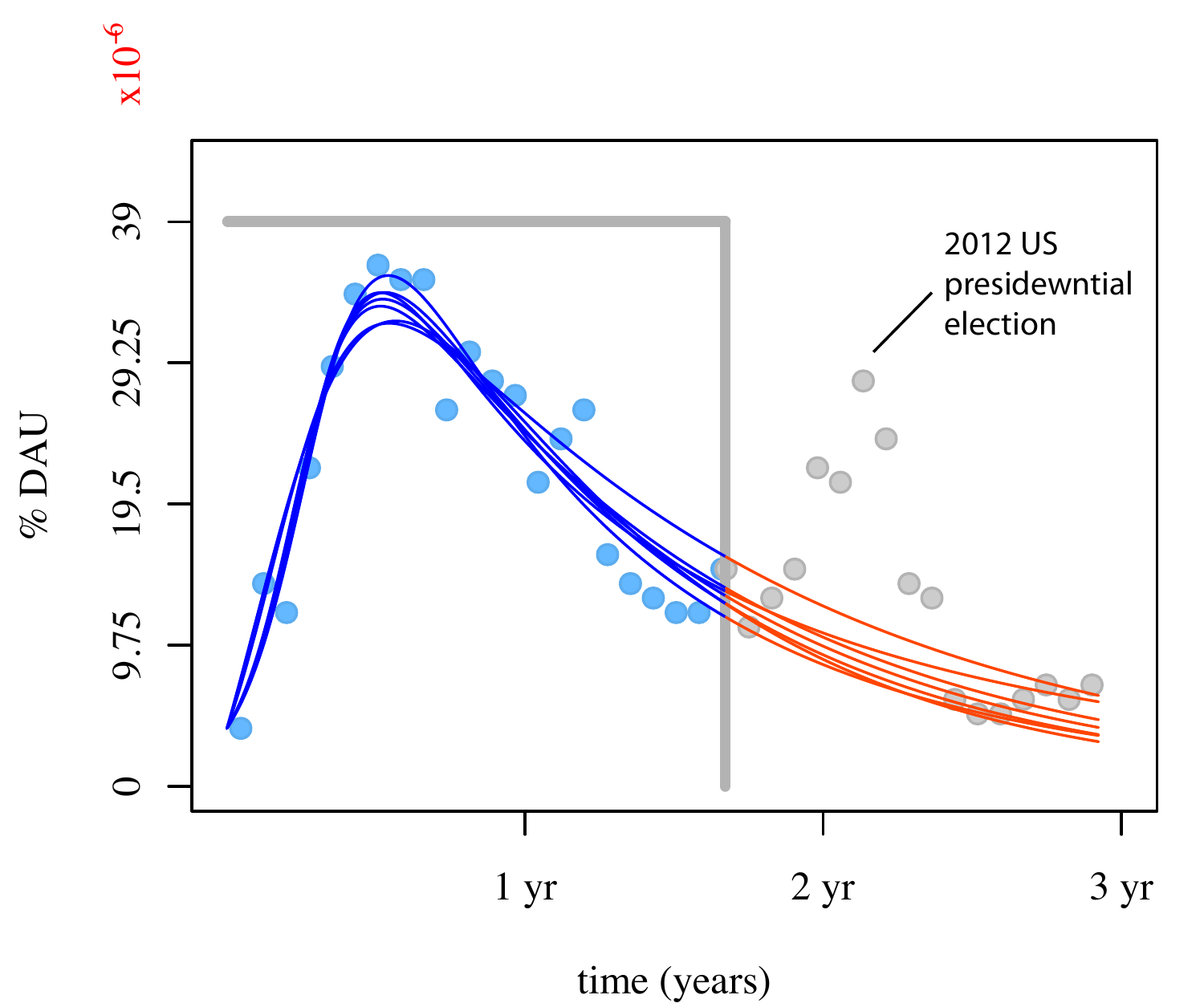}
}
\subfloat[][true.com\label{f:tru}]{
\includegraphics[width=1.6in,height=1.3in]{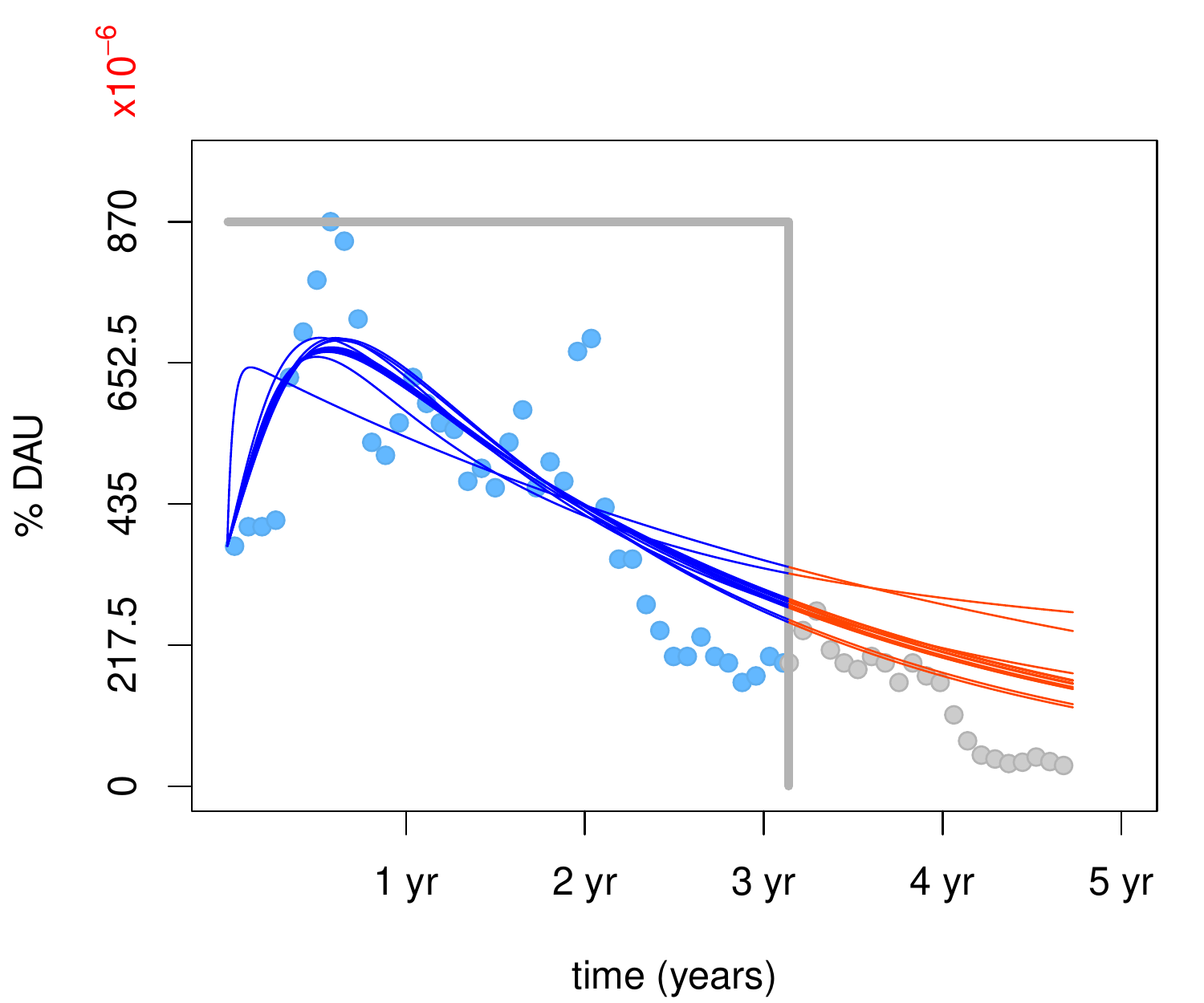}
}
\subfloat[][teapartypatriots.org\label{f:tep}]{
\includegraphics[width=1.6in,height=1.3in]{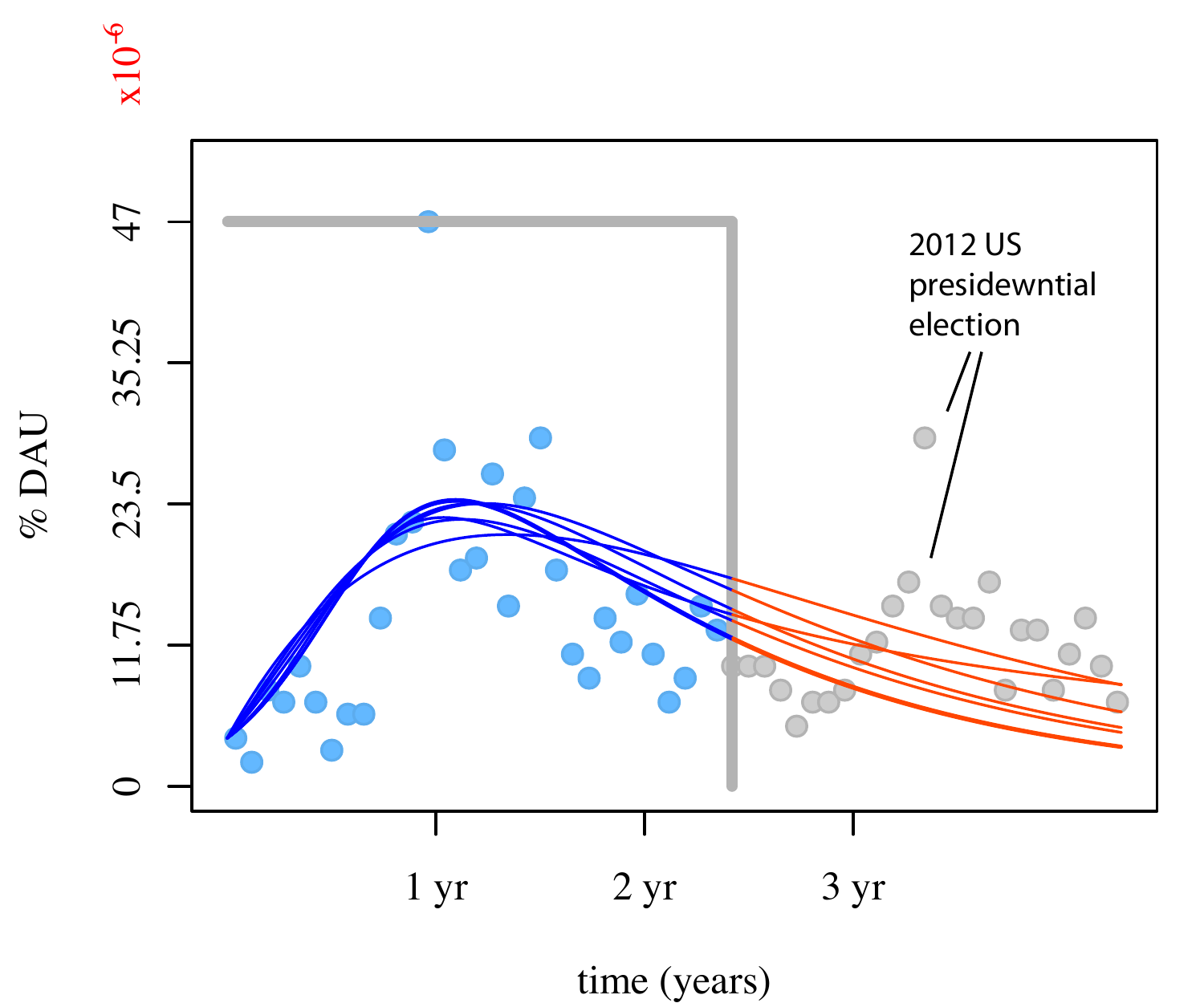}
}
\subfloat[][brandstack.com (before acquisition)\label{f:bra}]{
\includegraphics[width=1.6in,height=1.3in]{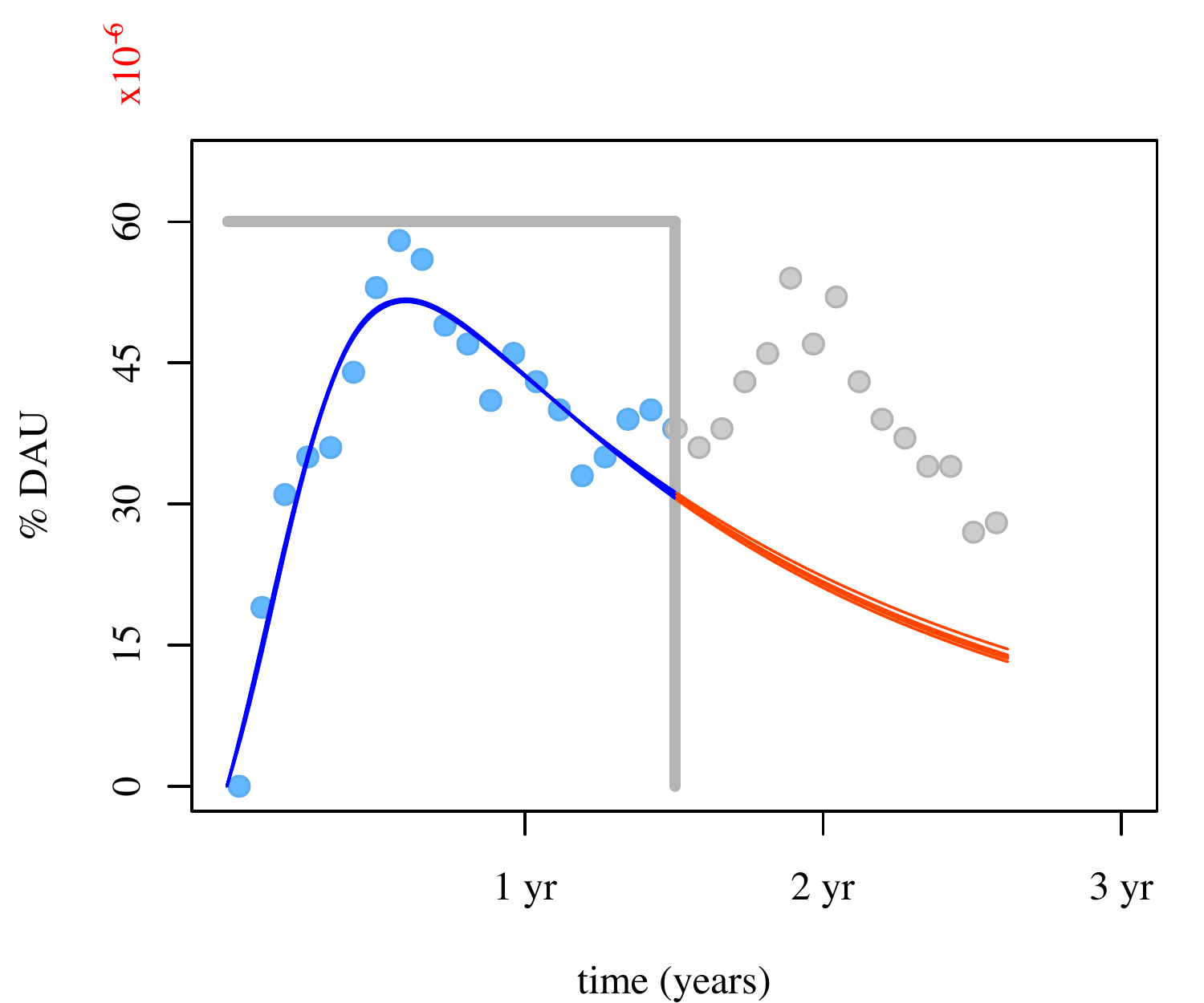}
}
\caption{{\bf (Unsustainable)} DAU time seires of websites automatically classified as unsustainable. Model fit (blue curve) over training samples and model selection samples (blue points) and model forecast (red line) over holdout data (gray points).\label{f:un}}
\end{figure*}

\begin{figure*}[!!ht]
\setcounter{subfigure}{0}
\centering
\subfloat[][Common self-sustaining + word-of-mouth DAU shape.\label{f:tsuwm}]{
\includegraphics[width=1.4in,height=1.2in]{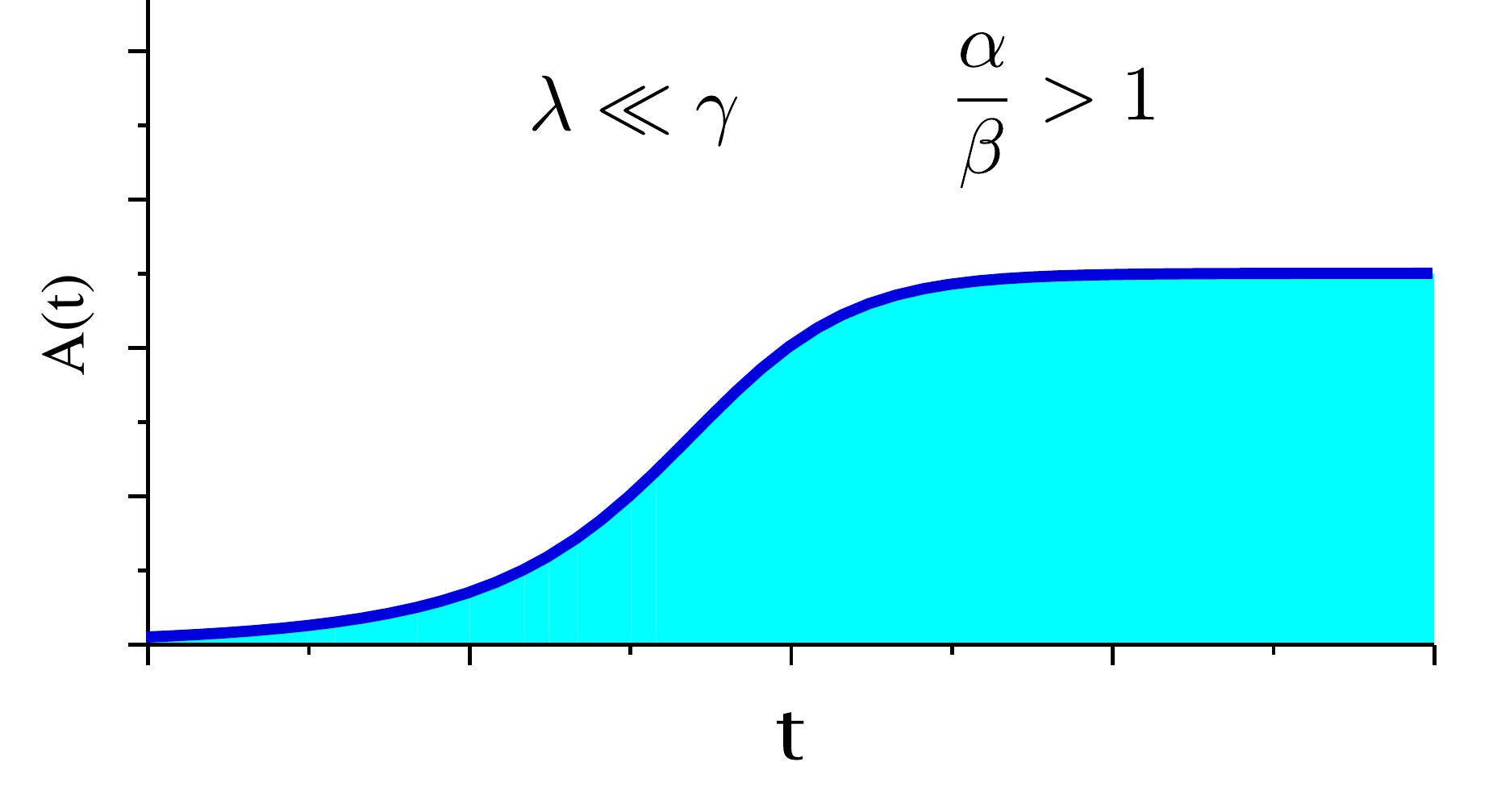}
}
\subfloat[][huffingtonpost.com\label{f:huf}]{
\includegraphics[width=1.4in,height=1.2in]{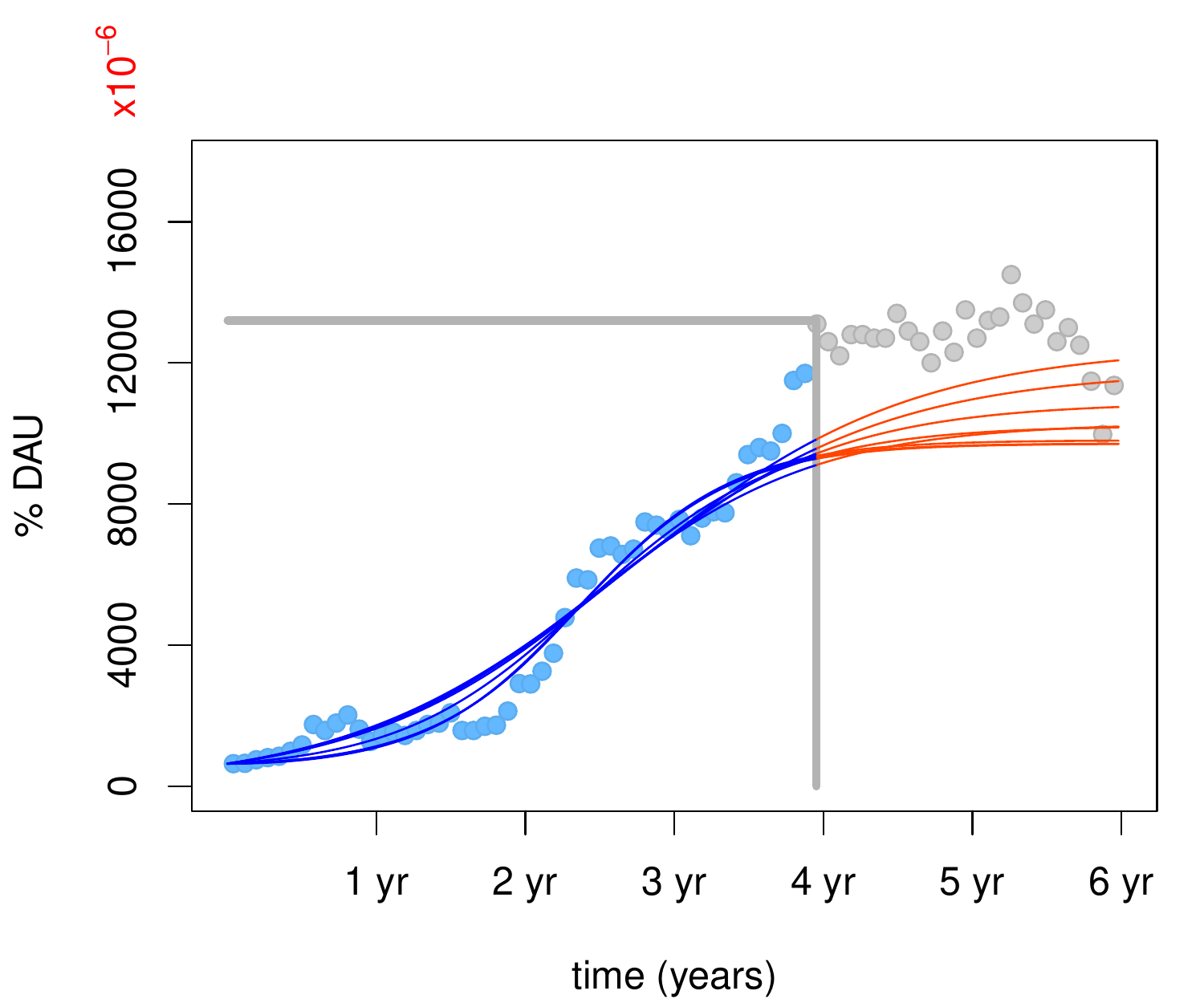}
}
\subfloat[][community.babycenter.com\label{f:rsus}]{
\includegraphics[width=1.4in,height=1.2in]{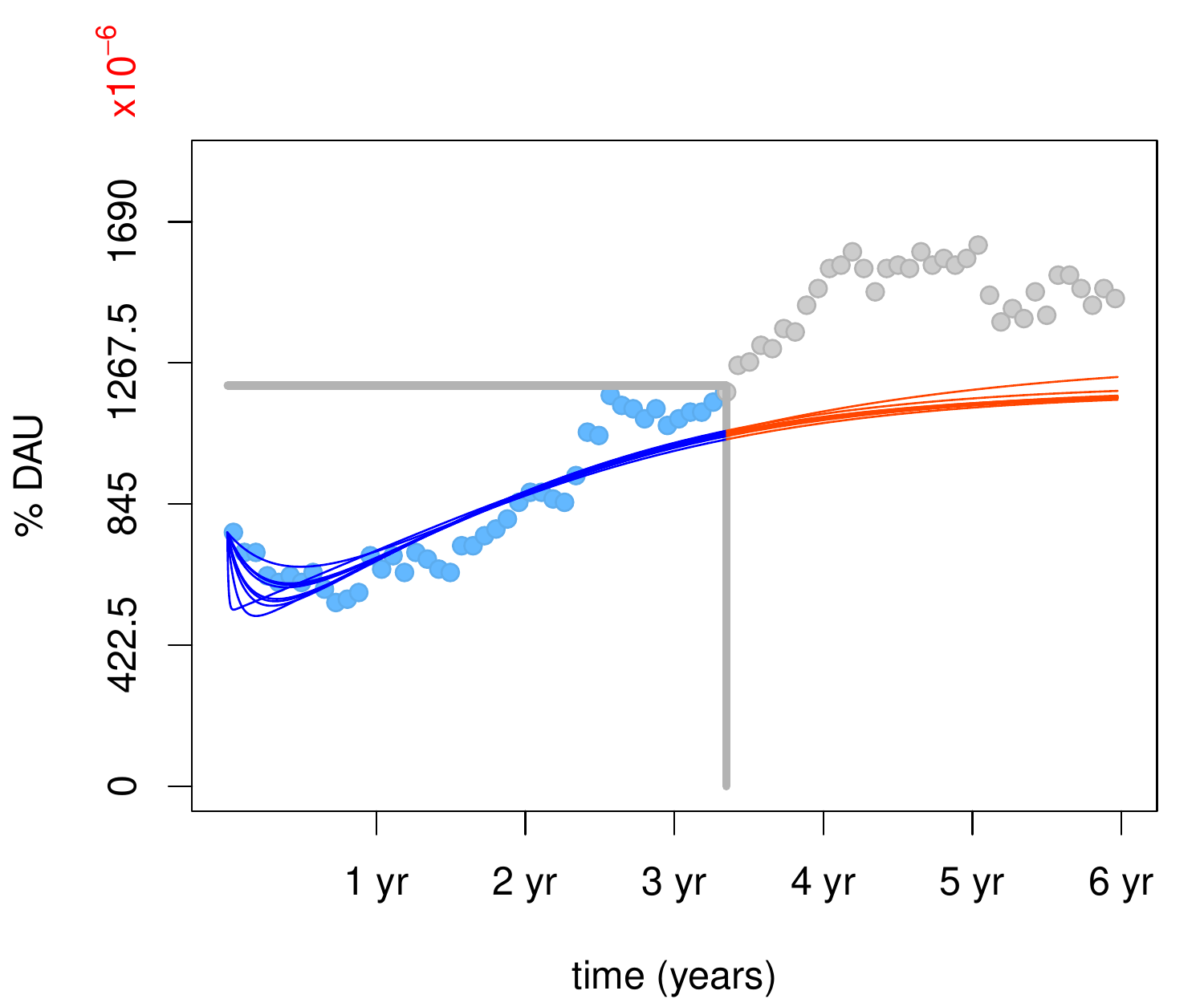}
}
\subfloat[][meetup.com\label{f:met}]{
\includegraphics[width=1.4in,height=1.2in]{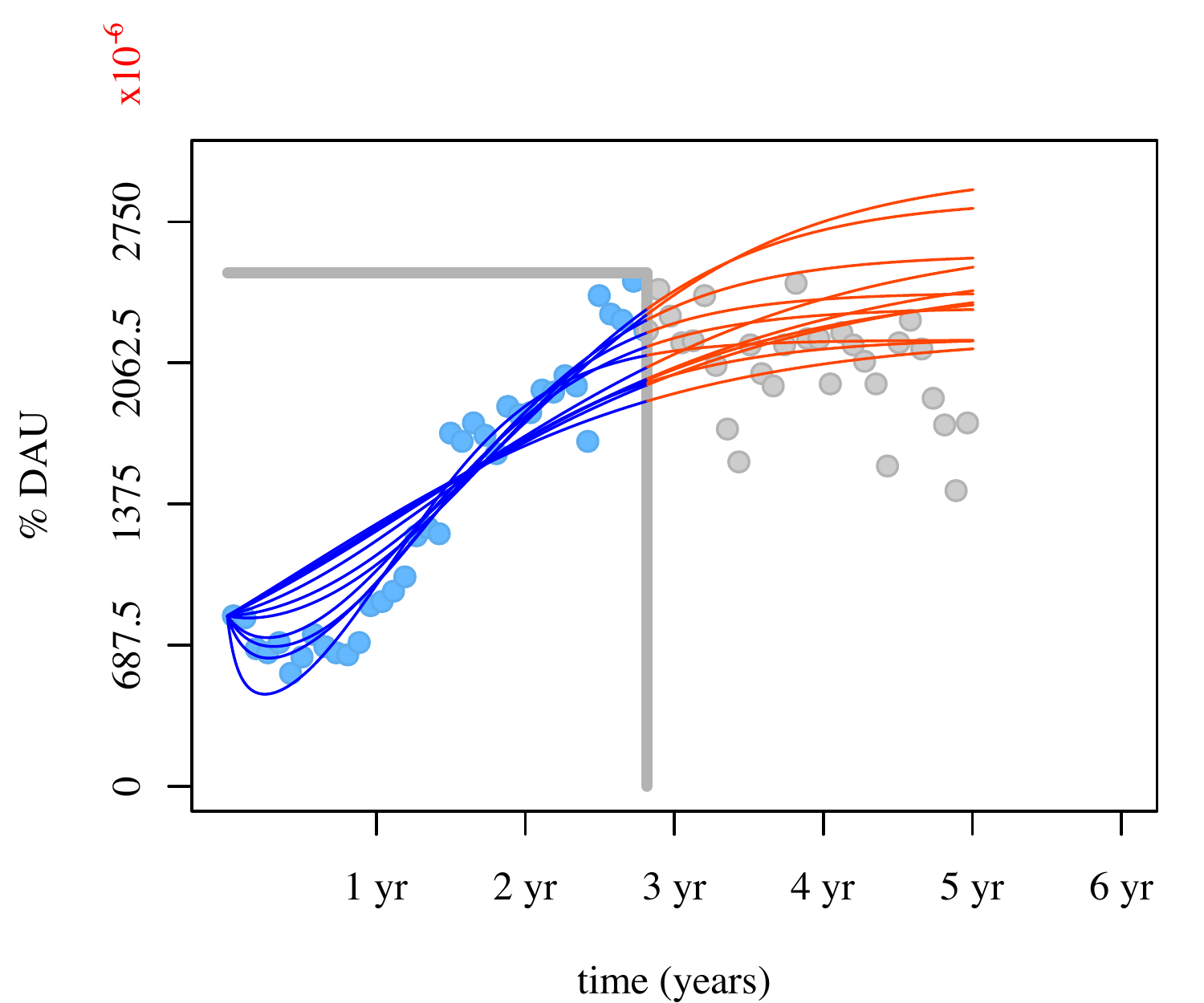}
}
\subfloat[][ashleymadison.com\label{f:ash}]{
\includegraphics[width=1.4in,height=1.2in]{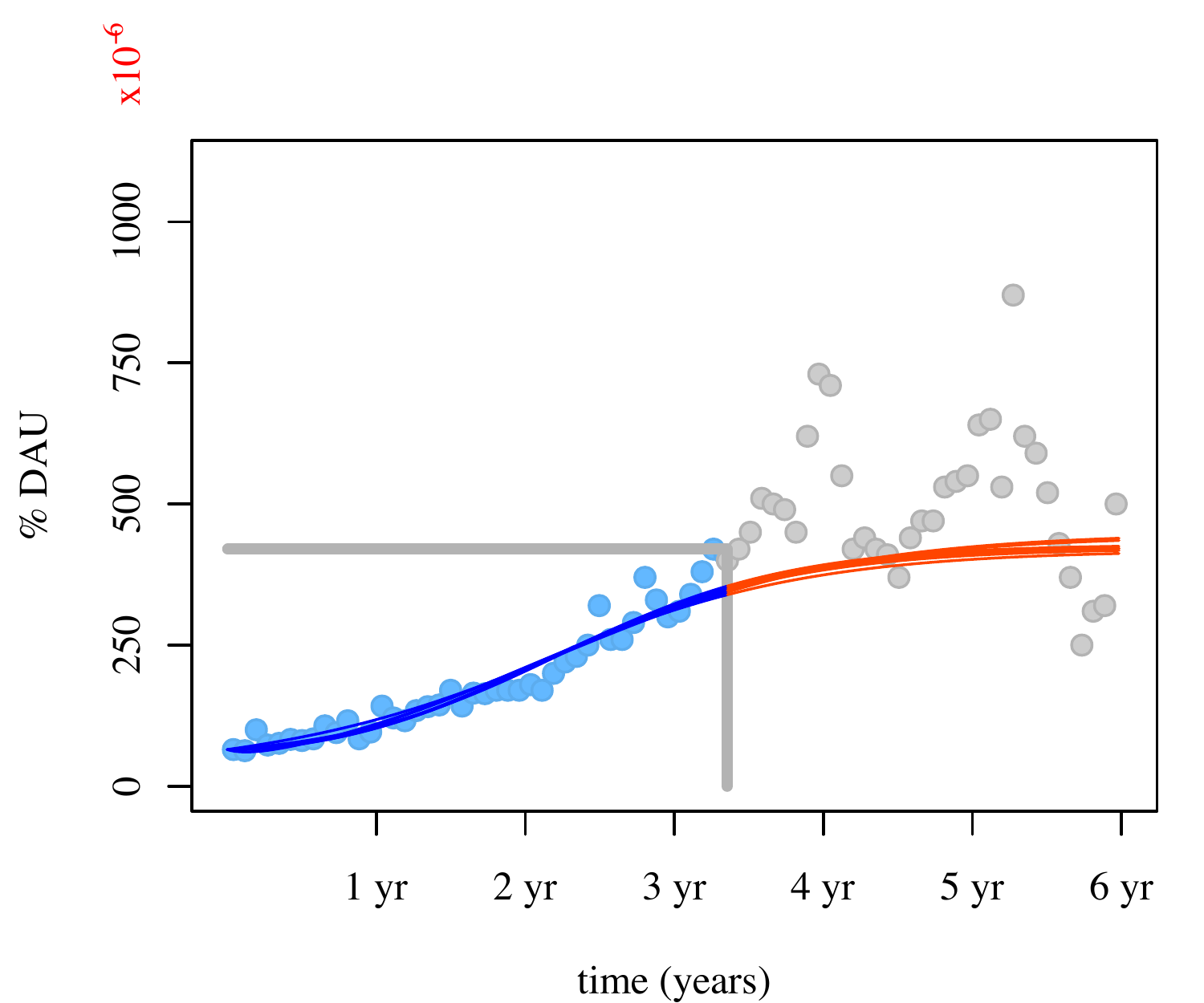}
}
\\
\subfloat[][facebook.com\label{f:fac}]{
\includegraphics[width=1.5in,height=1.2in]{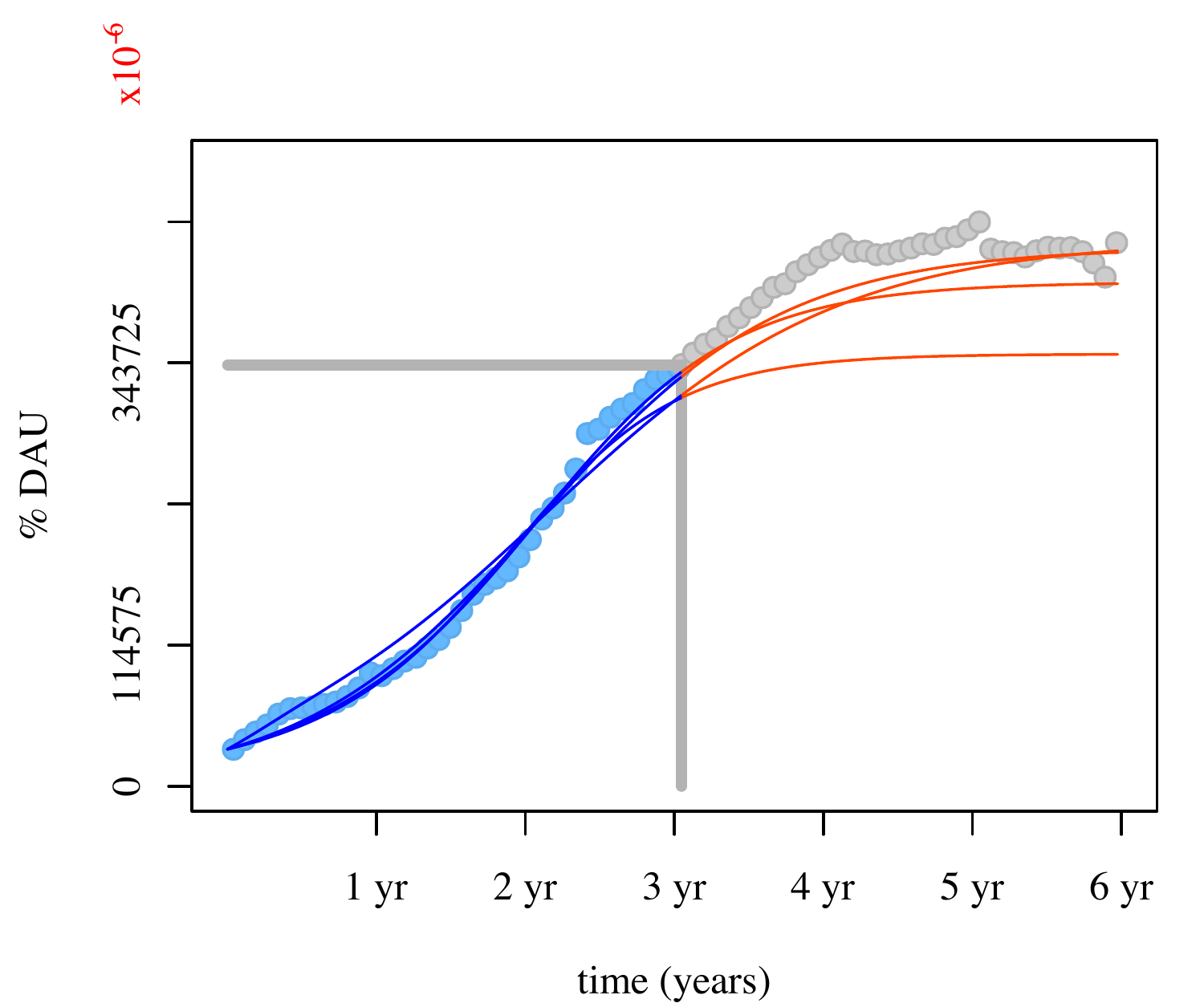}
}
\subfloat[][linkedin.com\label{f:lin}]{
\includegraphics[width=1.5in,height=1.2in]{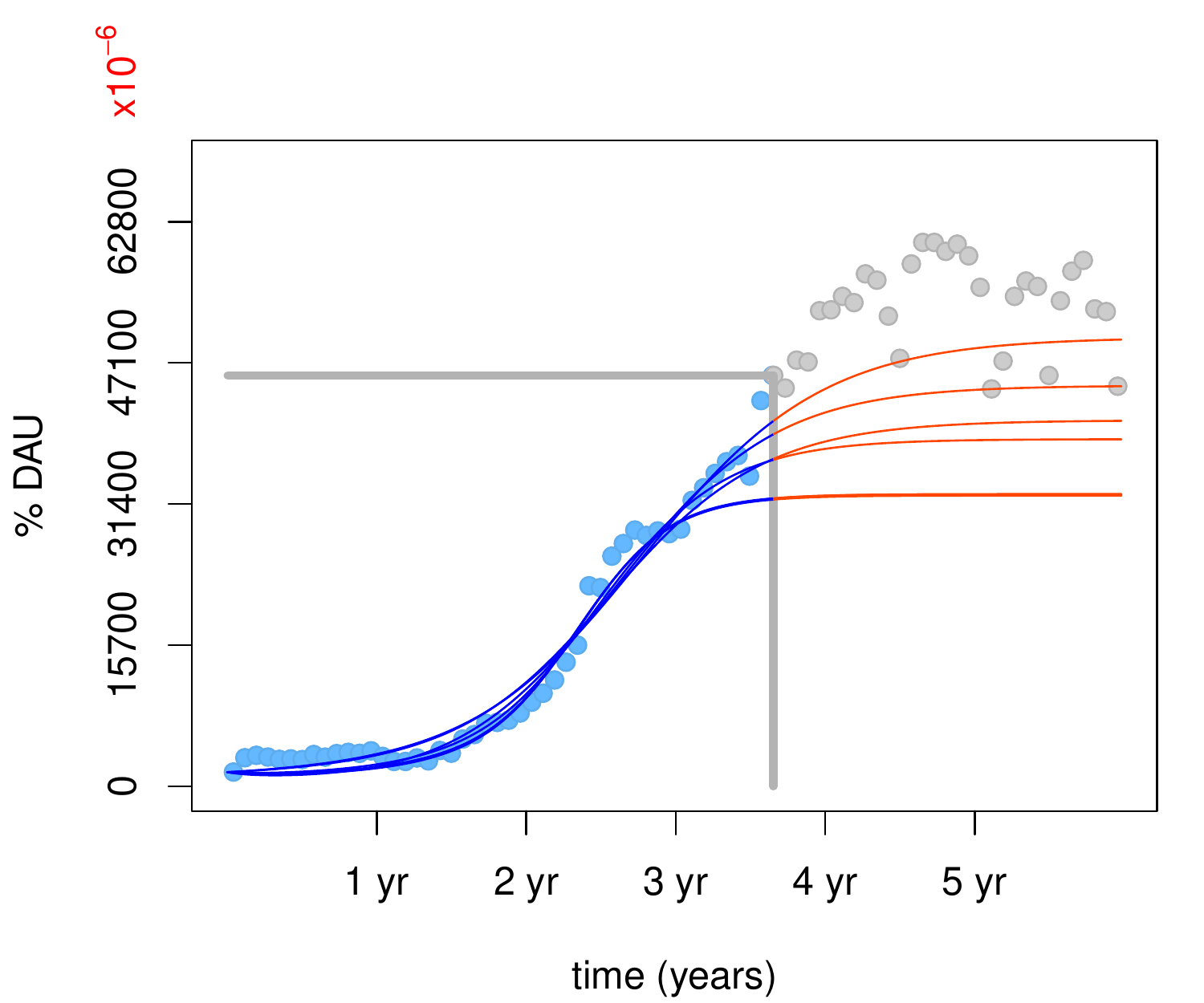}
}
\subfloat[][patientslikeme.com\label{f:pat}]{
\includegraphics[width=1.5in,height=1.2in]{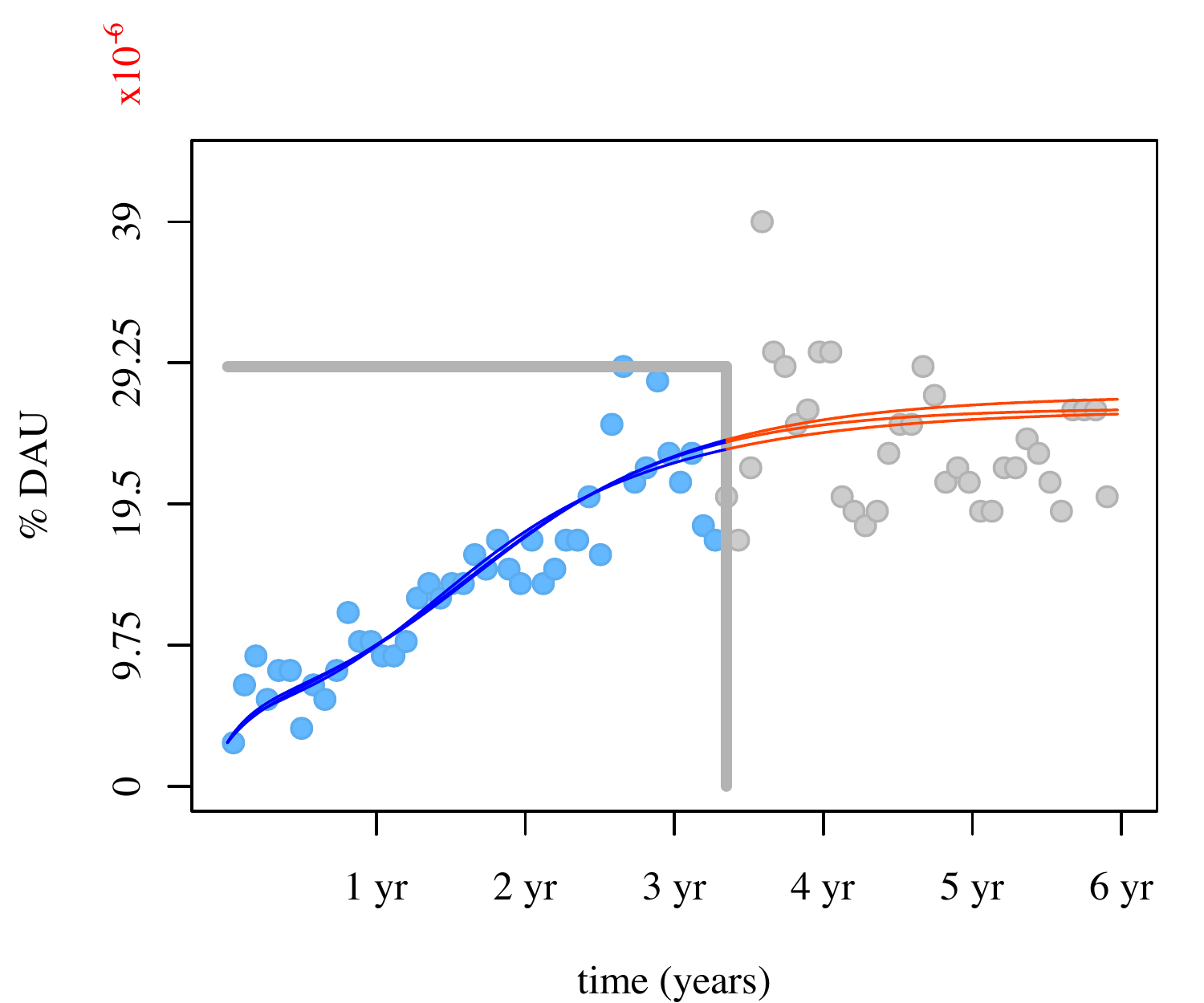}
}
\subfloat[][cafemom.com\label{f:caf}]{
\includegraphics[width=1.5in,height=1.2in]{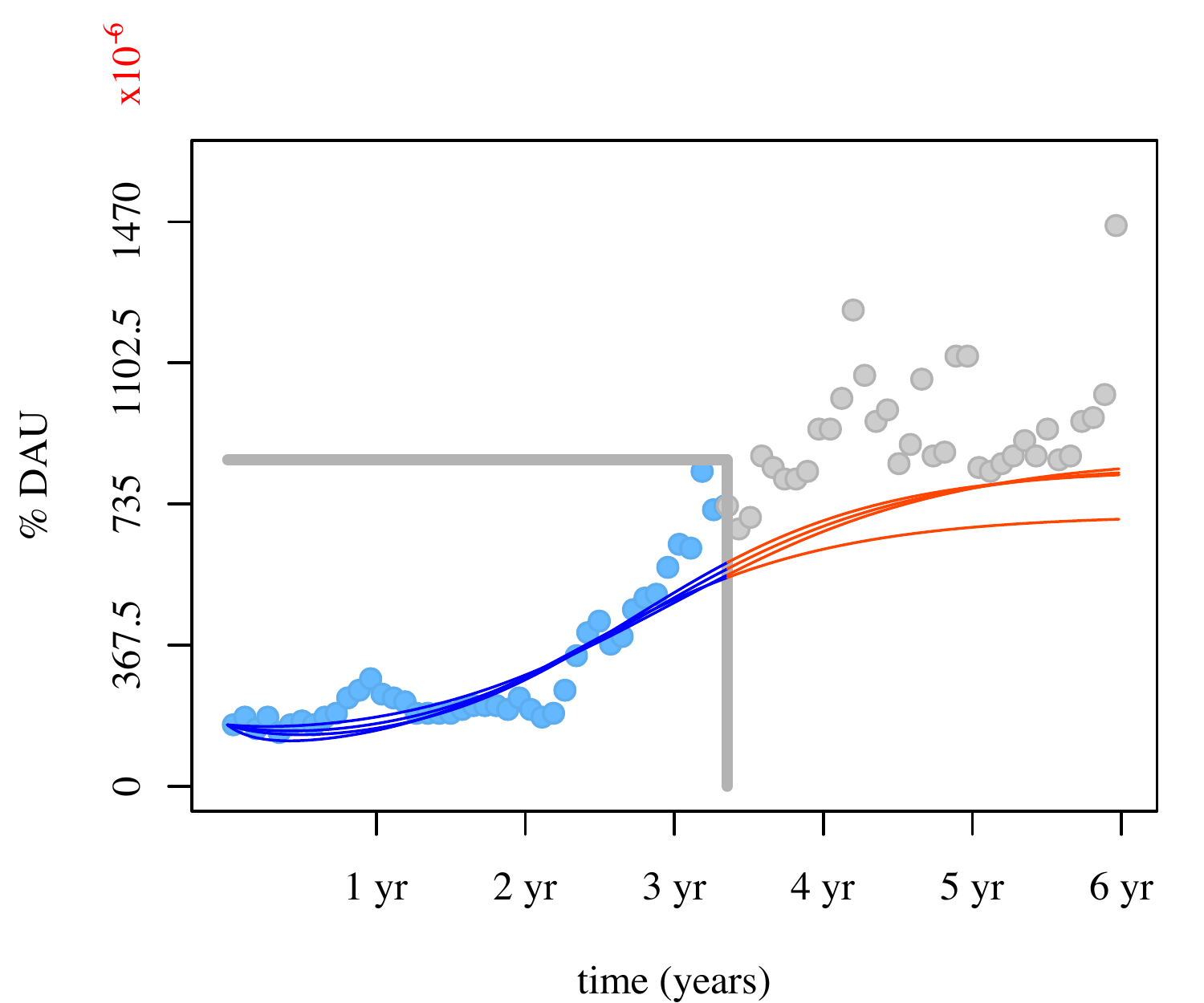}
}
\\
\subfloat[][marriedsecrets.com\label{f:mar}]{
\includegraphics[width=1.5in,height=1.2in]{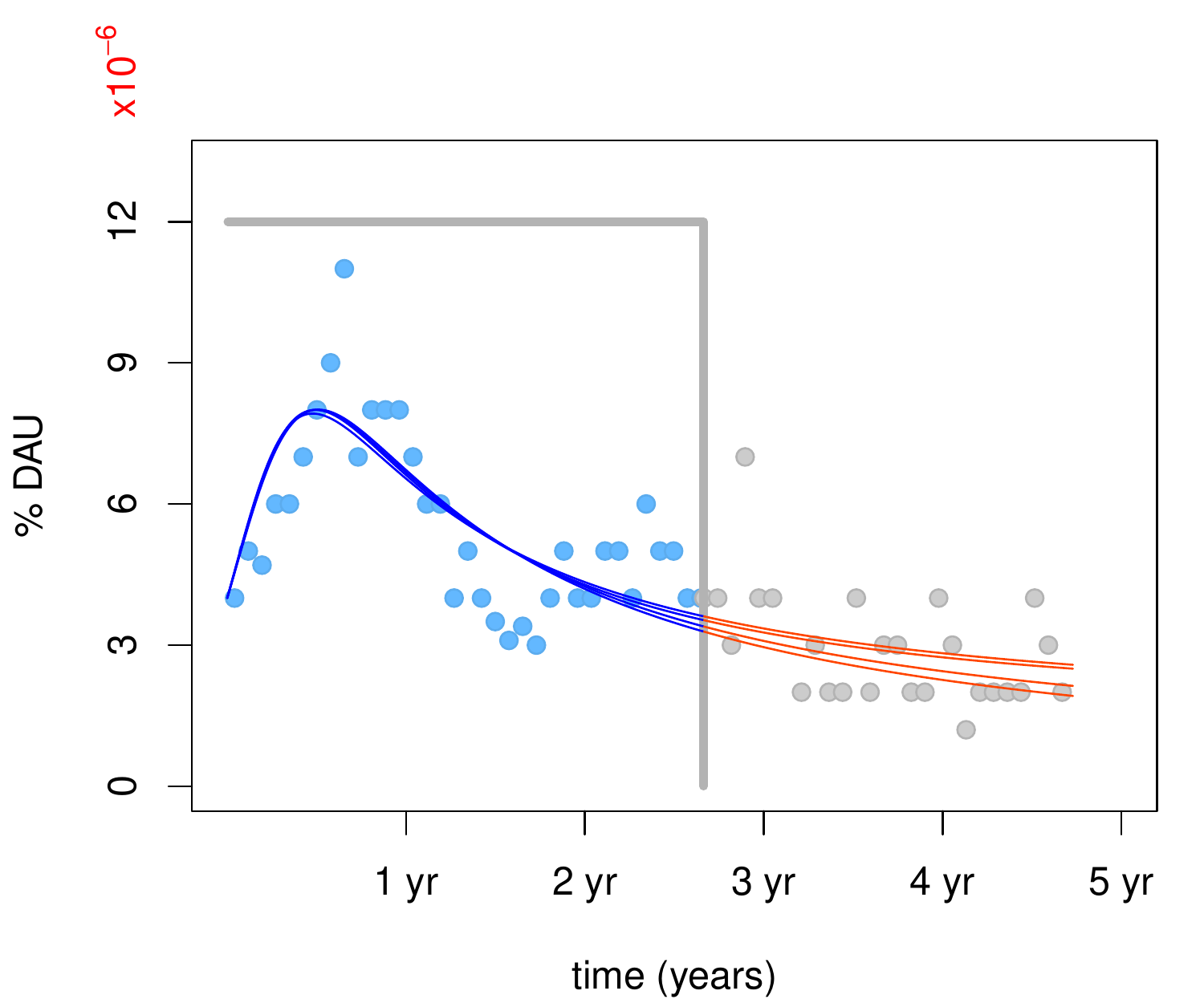}
}
\subfloat[][webchat.freenode.net\label{f:web}]{
\includegraphics[width=1.5in,height=1.2in]{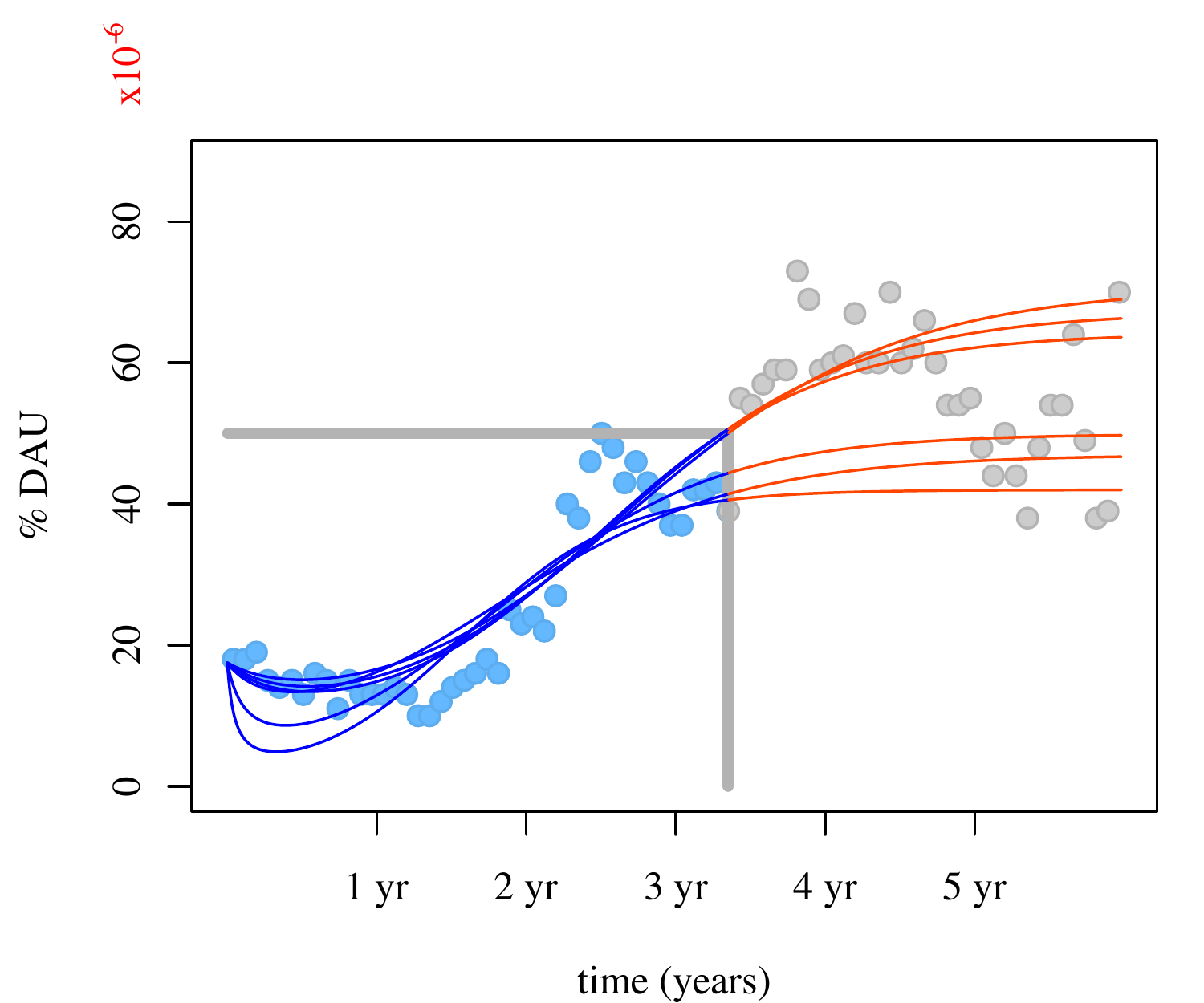}
}
\subfloat[][theblaze.com\label{f:bla}]{
\includegraphics[width=1.5in,height=1.2in]{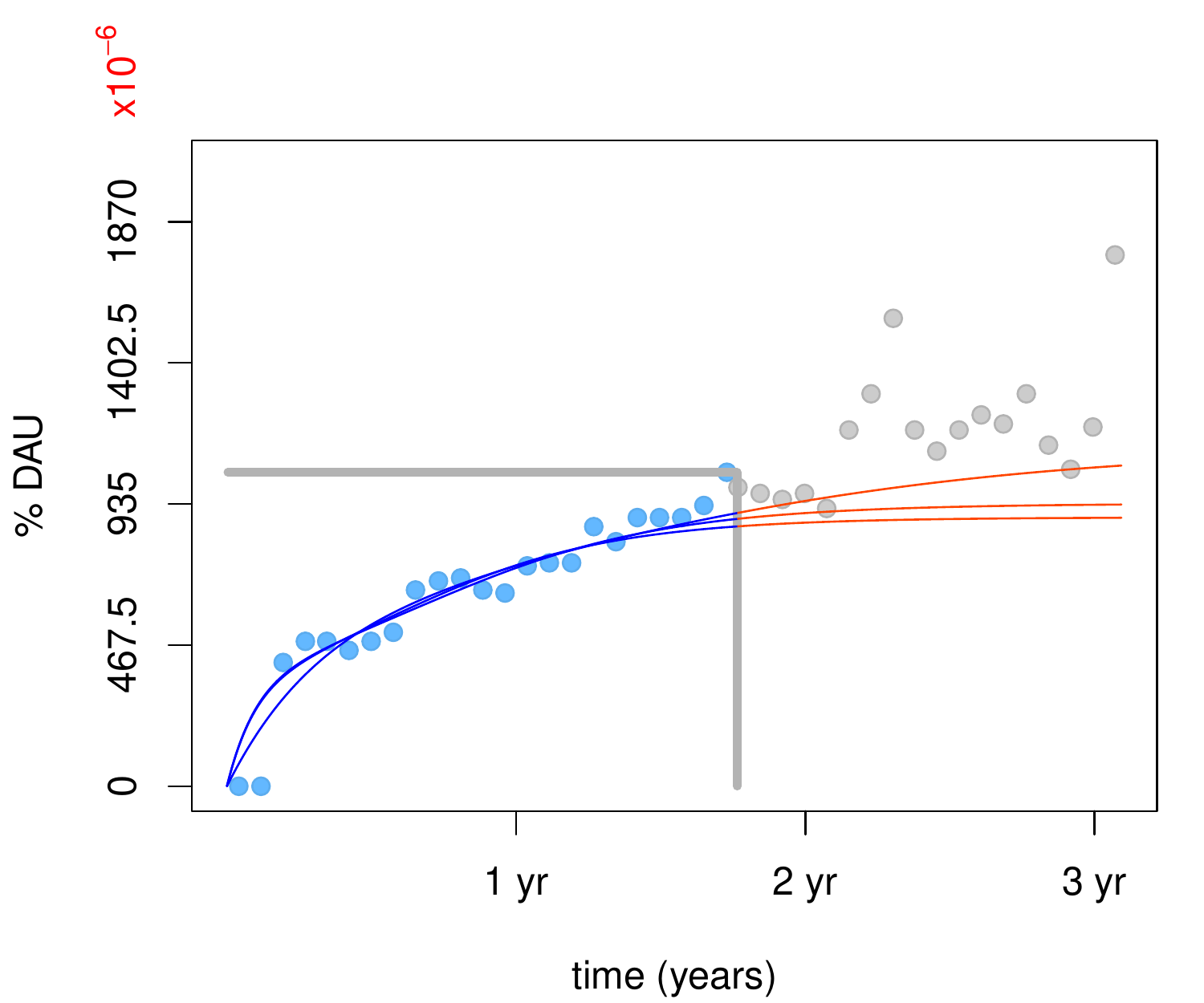}
}
\subfloat[][netflix.com\label{f:net}]{
\includegraphics[width=1.5in,height=1.2in]{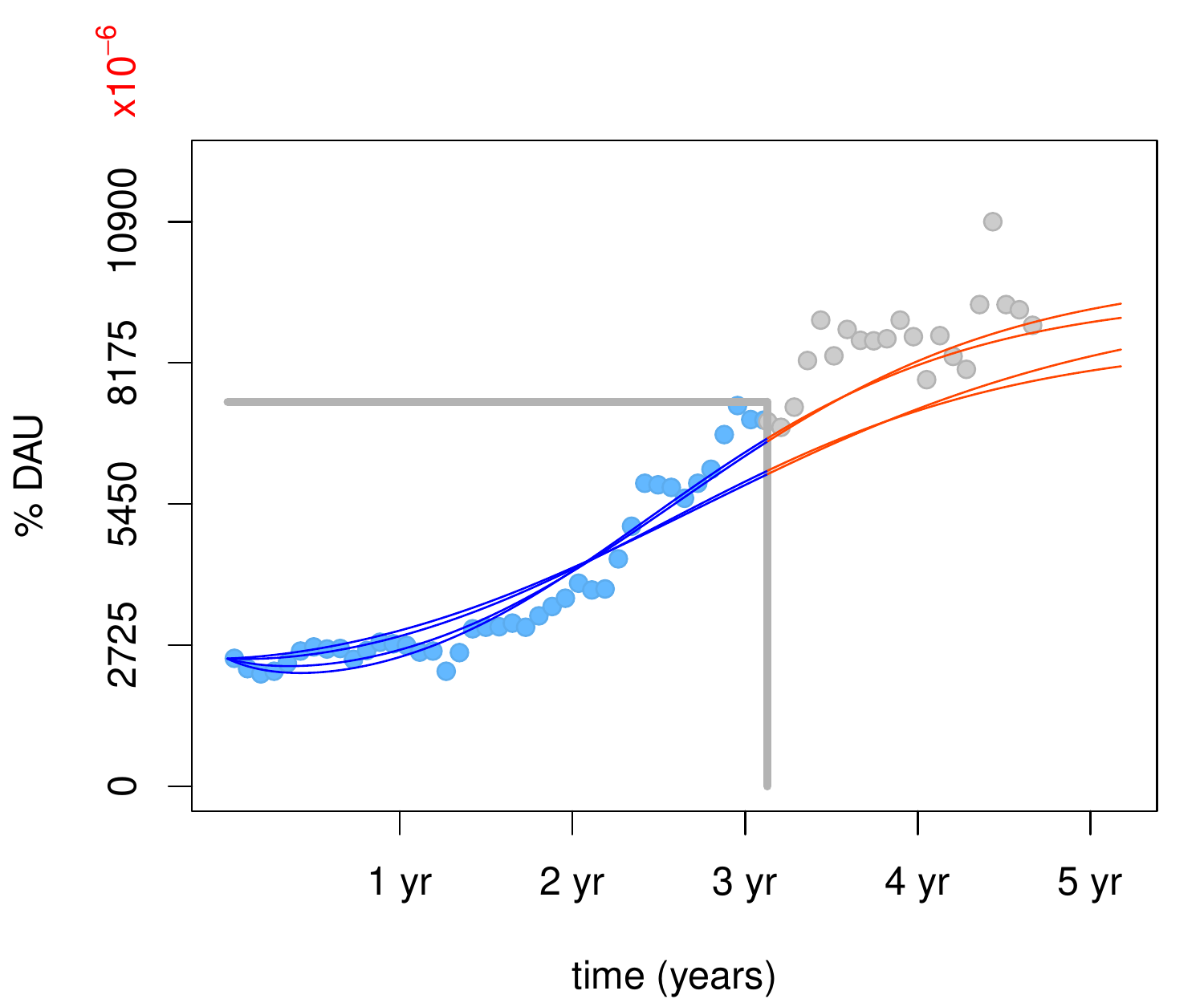}
}
\caption{{\bf (Self-sustaining)} DAU time seires of websites automatically classified as self-sustaining. Model fit (blue curve) over training samples and model selection samples (blue points) and model forecast (red line) over holdout data (gray points).
\label{f:su}
}
\end{figure*}

Our model has five parameters, namely, $\alpha$, $\beta$, $\lambda$, $\gamma$, and $C$.
Notwithstanding the five-dimensional parameter space, the overall DAU evolution shapes allowed by the model effectively has fewer degrees of freedom.
The reduction in degrees of freedom is due to the relationship  
between $\beta$ and $\alpha$ explored in Section~\ref{sec:lt}, the initial DAU growth differences between $\lambda$ and $\gamma$ explored in Section~\ref{sec:gr}, and the magnitude of $\lambda$ and $\gamma$ which helps dictate the DAU initial growth rate. Note that $C$ is just a resealing parameter and does not affect the shape of the DAU time series curve.
Figures~\subref*{f:tunwm},~\subref*{f:tunmk}, and~\subref*{f:tsuwm} present three of the most common DAU evolution shapes observed in our datasets.

The algorithm used to fit the model to the data works as follows.
Let $D$ be the set containing all datasets used in this study.
We find the an optimal parameter fit to a dataset $d \in D$ using the Levenberg-Marquardt algorithm~\cite{levenberg44}.
We use the first few years of DAUs data to train the model and the remaining $h$ years data as holdout data to evaluate the model predictions (part of $h$ is also user later in a model selection phase).
The Levenberg-Marquardt algorithm only finds a locally optimal solution starting from an initial parameter guess $m_0 = (\alpha_0,\beta_0,\lambda_0,\gamma_0,C_0) $.
Hence, the initial guess $m_0$ may significantly influence the output of the algorithm.
To make the algorithm fully automatic and robust we need to find a principled way to provide this initial guess using our datasets.

We provide the initial parameter guess by feeding the Levenberg-Marquardt algorithm with other parameter examples obtained in our datasets.
It is reasonable to assume that the DAU time series of a given dataset $d \in D$ is similar to the DAU time series of some other datasets in $D$.
If we knew the best parameter fit of a dataset with similar DAU time series as $d$ we could then use it to initialize the Levenberg-Marquardt algorithm.
But as we don't know which datasets in $D \backslash \{d\}$ ($D \backslash \{d\}$ is the set of all datasets excluding $d$) have similar DAU,
we test all possible datasets as follows.
Let $\LM(d,m_0)$ be the Levenberg-Marquardt fitted parameters using initial guess $m_0$.
Let $f_{d^\prime}$ be the best currently known fit to dataset $d^\prime \in D \backslash \{d\}$.
At this stage our algorithm produces a set of candidate fitted parameters $\cC = \{\LM(d,f_{d^\prime}) : d^\prime \in D\backslash \{d\}\}$. 

Instead of selecting the best fitted parameters from $\cC$, we make our parameter selection more robust by selecting at least three fitted parameters from $\cC$.
We test the quality of the fitted parameters through a model selection phase.
In the model selection phase we use the first three to six months of DAU data in the holdout data (comprising of up to $10\%$ of the total data) 
to assign an $L_2$ error to the predictions of the model with each fitted parameters in $\cC$.
Using these errors, together with the actual parameter values, we cluster the elements of $\cC$ using $k$-medoids clustering (using the R package {\em pamk} and the Calinski-Harabasz~\cite{calinski1974dendrite} criteria to automatically select the number of clusters $k$).
We choose to use $k$-medoids instead of the widely used $k$-means because $k$-medoids is likely more robust to noise and outliers than $k$-means, in the same manner that the median of a set of measurements is more robust to noise and outliers than their mean.
The output of our algorithm is then the $k$-medoid cluster $c \subseteq \cC$ that has at least three elements and the smallest average error.

More precisely, our algorithm reports $c$ and the medoid of $c$.
The medoid of $c$ is the vector of parameters that best represents all vectors of parameters in $c$. 
The above procedure must be bootstrapped by manually assigning parameter fits to the datasets. 
But once we automatically obtain the medoid parameter fit of $d$ we can replace its manually initialized value with the automatic one and restart the process.
The R source code of our algorithm is freely available online to be tested on other datasets\footnote{\scriptsize\url{http://www.cs.cmu.edu/~ribeiro/Ribeiro_AIU.zip}}.

\subsection{Model fitting results}
In what follows we present the model fitting results of our algorithm (more precisely the medoid value) and later contrasts them against the real DAU time series.
The magnitude of the fitted model parameters should be compared between websites as they need to be rescaled by $C$, which varies from dataset to dataset.
For instance, the value of $C$ for Facebook is much larger than that of any other website in our dataset.
\\

\noindent
({\bf Long-term DAU sustainability})
Figure~\subref*{f:ab} shows the medoids of the fitted parameters $\alpha$ and $\beta$ for each of the twenty-two websites.
The light blue area in Figure~\subref*{f:ab} indicates the long term unsustainable region where $\beta/\alpha \geq 1$. 
The DAU of websites in this region is predicted to die in the long run.
In our plots whenever the parameter values are smaller than $10^{-4}$ we declare the parameter as N\!/\!A indicating that the required precision may be beyond the numerical capabilities of our algorithm.
Interestingly, websites that are known to have a sustained loyal fan base such as Facebook, LinkedIn, Salon, DailyCaller, and TheHuffingtonPost, are unmistakably  in the {\em self-sustaining} area of the $\alpha$ and $\beta$ parameter space.

On the other hand, websites of social movements such as the Tea Party websites teapartypatriots.org and teapartynation.com, along with OccupyWallSt.org rest squarely on the ``unsustainable'' area.
Internet fads such as 12seconds.tv and  also belong to the unsustainable area.
Other websites such as Flixster.com, FormSpring.me, and MarriedSecrets.com are near the border between self-sustaining and unsustainable.
MarriedSecrets.com and AshleyMadison.com, unlike a regular dating websites, are designed to support extra-marital affairs of both genders.
We choose these dating websites instead of more traditional dating websites such as OkCupid.com and Match.com as we believe that people looking for extra-marital affairs are not likely to quit the website once they find a partner, thus making their behavior more similar to our model.\\

\noindent
({\bf Signatures of growth}) Fig.~\subref*{f:mw} shows the medoids of the fitted parameters $\gamma$ and $\lambda$ and the respective website.
Observe that most websites are classified as having a significant word-of-mouth component with a comparatively small media \& marketing push.
Few websites such as OccupyWallSt.org and community.babycenter.com are classified as having both strong word-of-mouth and media \& marketing exposure.
Only two websites were classified as having predominantly media and marketing campaigns: ebay.com and theblaze.com.
Indeed, eBay has strong visibility on search engines for searches related to electronics and hard-to-find consumer products,
to which its ``marketing'' visibility could be attributed.
The TheBlaze.com result is more mysterious.

TheBlaze.com is an online conservative news venture run by conservative pundit Glenn Beck~\cite{TheBlaze}. 
Contrasting the collected DAU time series of TheBlaze.com in Fig.~\subref*{f:bla} with the shapes of strong media \& marketing intensive DAU signatures in Fig.~\ref{f:ado}(a) and word-of-mouth intensive DAU signatures in Fig.~\ref{f:ado}(b) it is clear that TheBlaze.com is correctly classified as media \& marketing intensive growth with little word-of-mouth growth.
Unfortunately, it is unclear why TheBlaze.com presents an intensive media \& marketing growth signature while similar political news websites such as DailyCaller.com and TheHuffingtonPost.com have the opposite classification (little media \& marketing growth with strong word-of-mouth growth).

While our model is designed to capture just the basic mechanism that drives the DAU in the absence of sudden events that may happen during the website's lifetime -- e.g., deep website redesigns, extra marketing and media exposures, and arrivals of new strong competing websites vying for the attention of the website members --
in what follows we observe that even in the presence of unpredictable DAU bumps and spikes our model is able to capture the DAU time series and predict its long-term trend.

\subsection{DAU fit \& long-term predictions}
Here we present the data fit and the predictions made by our model.
First, however, we introduce the most common combinations of the DAU signatures of self-sustaining v.s.\ unsustainable and word-of-mouth v.s.\ marketing \& media intensive growths in our datasets.
Figs.~\subref*{f:tunwm} and~\subref*{f:tunmk} use our model to show the most common DAU shapes of unsustainable websites observed in our datasets.
Fig.~\subref*{f:tsuwm} uses our model to show a common word-of-mouth intensive DAU shape of sustainable websites.
Few websites are classified as sustainable with media \& marketing intensive growth (more specifically just the TheBlaze.com and BrandStack.com websites).

One of our goals in this section is the challenging task of presenting multi-year DAU predictions into the future based on the various vectors of parameters obtained by our algorithm from the training  and model selection data.
It is worth noting that parameterizing the model using only DAU information has its limitations.
It is only possible to classify a website as unsustainable if the website shows signs of decreasing DAU activity.
Interestingly, however, is that not all websites showing signs of decreasing DAU activity are classified as unsustainable, 
as exemplified by the website MarriedSecrets.com.

In Figs.~\ref{f:un} and~\ref{f:su} the DAU time series is shown as points. 
The horizontal axis shows the years since the first recorded website traffic activity starting from June 2007.
The vertical axis shows the DAU as a percentage of the active Internet population on any particular day (marked as \%DAU).
We thin the number of points displayed in the plots just to avoid clutter.
The blue points show the DAU data used in the training and model selection phases.
The gray points shows the holdout DAU data that our model needs to predict.
The gray box shows the maximum vertical (\%DAU) and horizontal (time) values of the combined training and model selection data.
The blue curves show the fit of the best collection of model parameters output by our algorithm (all the parameters in the $k$-medoid cluster $c$).
\\

\noindent
({\bf Unsustainable websites)} Fig.~\ref{f:un} shows the results of the twelve websites automatically classified as unsustainable by our algorithm. 
We observe that the model fits the data reasonably well, except for a few ``DAU bumps'', possibly the product of website changes (which may temporarily change the model parameters) or media \& marketing investments.
In general, however, our model is able predict the overall website trend years in advance.
Three anomalies are the overestimation of the DAU of True.com and Adaptu.com.
Per our Tech Report~\cite{techreport} the linear decay of Adaptu.com's DAU shows the characteristic signature of a strong competitor (or even multiple strong competitors), probably Mint.com~\cite{deadpool}.
Our fitting algorithm also tends to output some parameter assignments for Ruelala.com as self-sustaining. 
BrandStack.com was acquired in late 2011 by DesignCrowd.com and has since changed its  url and thus we truncate its data at the time of acquisition.
The predicted DAU trend of BrandStack.com seems accurate if the website was allowed to continue its trajectory.
The predictions of Flixter.com, OccupyWallSt.org, 12seconds.tv, FormSpring.me, TeaPartyNation.com, True.com, and TeaPartyPatriots.org are remarkably accurate.
Interestingly, not even the renewed interest in Tea Party during the  2012 U.S.\ presidential election cycle (marked in the graph) was able to disturb the long-term trajectory of TeaPartyNation.com and TeaPartyPatriots.org towards what appears to be a negligible (zero) DAU.
\\

\noindent
({\bf Self-sustaining websites)} Fig.~\ref{f:su} shows the DAU time series (points) of twelve websites automatically in our dataset that are classified as self-sustaining by our algorithm.
One of the most remarkable predictions of our model is the long term stabilization of the DAU of self-sustaining websites.
The datasets (TheHuffingtonPost.com, Facebook.com, among others) overwhelmingly confirm this prediction. 
While websites keep constantly changing to broaden their audience (which could potentially lead to an increase of $C$ in our model), it seems that once the website audience is determined the website \%DAU trajectory towards $C(1 - \beta/\alpha)$ is set.
One of the most remarkable examples of this observation is Facebook.com. 
Three out of the four outputs of our algorithm predict  three years into the future -- and with surprising accuracy -- Facebook's final DAU size and time series. The training + model selection data has Facebook's DAU reaching up to 37\% of the active Internet population and our model is able to extrapolate that Facebook would eventually reach about 45\% of the \%DAU and then stabilize.

The DAU fit and predictions of our model for AshleyMadison.com, Facebook.com, PatientsLikeMe.com, CafeMom.com, MarriedSecrets.com, Webchat.Freenode.net, and Netflix.com worked remarkably well. Both Meetup.com and
Despite the high variability of predictions at Meetup.com and  Webchat.Freenode.net, these predictions seem to agree well with the DAU data (which also present high variability).

Curiously, while the curve fit for TheHuffingtonPost.com is remarkably accurate, the model has a strong tendency to stabilize the DAU while the true data shows a growing DAU until stabilization.
The reason behind this apparent mismatch may be the following. At the beginning of the TheHuffingtonPost.com life its DAU growth was mostly due to word-of-mouth, which is well captured in the model parameters over the training data.
However, reports show that somewhere between early 2011~\cite{HuffPostGN} and late 2013~\cite{HuffpostGN2} the TheHuffingtonPost.com started to be (significantly) indexed by the Google News service, bringing a large amount of ``unpredictable'' media \& marketing growth to the TheHuffingtonPost DAU, thus the DAU underestimation of our model.
Similar to the TheHuffingtonPost.com the DAU of LinkedIn.com and community.babycenter.com are also shown to be self-sustaining (i.e., the DAU stabilizes) but the final DAU is underestimated. Unfortunately the reasons behind these two underestimations are unclear.

The website TheBlaze.com seems to have also slightly deviated from its predicted DAU. In the first 112 days into the prediction the website DAU follows the model remarkably well. Later, however, TheBlaze.com experiences two unpredictable DAU spikes and, unfortunately, only time will tell if the model prediction is indeed showing the long-term DAU plateau of the website.

\section{Conclusions} \label{sec:con}

Our study sheds light on the mechanisms of growth and member activity and inactivity of membership-based websites.
Through reaction, diffusion, and decay processes we model the dynamics of website member activity, inactivity, and growth, mostly focusing on the DAU (daily number of active users) metric.
We showed that our model predicts two general DAU signatures of growth (media \& marketing versus word-of-mouth) and two DAU signatures of long term stability (self-sustaining versus unsustainable).
We proposed an algorithm to fit the model parameters to real-world DAU time series data.
Finally, from the DAU time series of twenty-two websites we show that our model not only fits well the DAU data but can also predict its future evolution.

This work makes a positive step not only  towards modeling the dynamics of websites but possibly also towards modeling
a broad range of  dynamics of societal movements, such as the activity and growth of grass-root organizations.
There is, however, much left to do to tailor the above reaction-diffusion-decay dynamics to specific types of websites
and changing environments (e.g., new technologies and competition).

\section{Acknowledgments}
{
  \fontencoding{T1}
  \fontfamily{times}
  \fontsize{8}{8}
  \selectfont

The author would like to thank the members of the MoBS lab at Northeastern University for their kind hospitality and insightful discussions, Sharon Goldberg for pointing out some missing related work, Don Towsley for encouraging the generalization of an earlier version of this work,  and  A.\ Clauset for bringing to the author's attention his study on online multiplayer networks. 
This work was partially done while the author was at the University of Massachusetts Amherst. This work was supported by NSF grant CNS-1065133 and ARL Cooperative Agreement W911NF-09-2-0053. 
The views and conclusions contained in this document are those of the author and should not be interpreted as
representing the official policies, either expressed or implied of the NSF, ARL, or the U.S. Government. The U.S. Government is authorized to reproduce and distribute reprints for Government purposes notwithstanding any copyright notation hereon.
}

\section{Appendix}

\subsection{Simpler model oversimplifies problem}
\begin{figure}[!tb]
\begin{center}
\subfloat[][Incorrect simpler model prediction\label{f:simp}]{
\includegraphics[width=1.4in,height=1.2in]{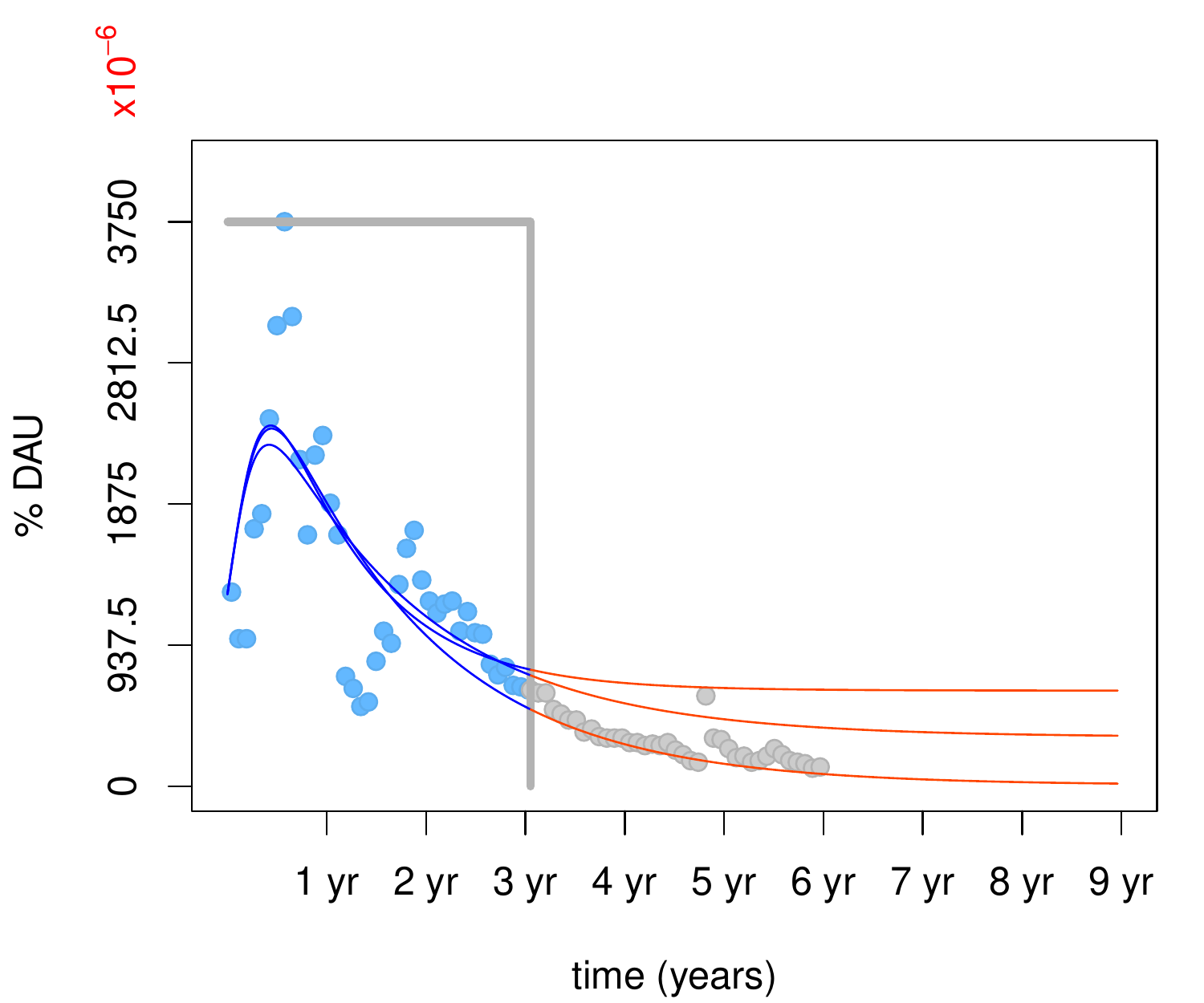}
}
\hspace{10pt}
\subfloat[][Complete model prediction\label{f:comp}]{
\includegraphics[width=1.4in,height=1.2in]{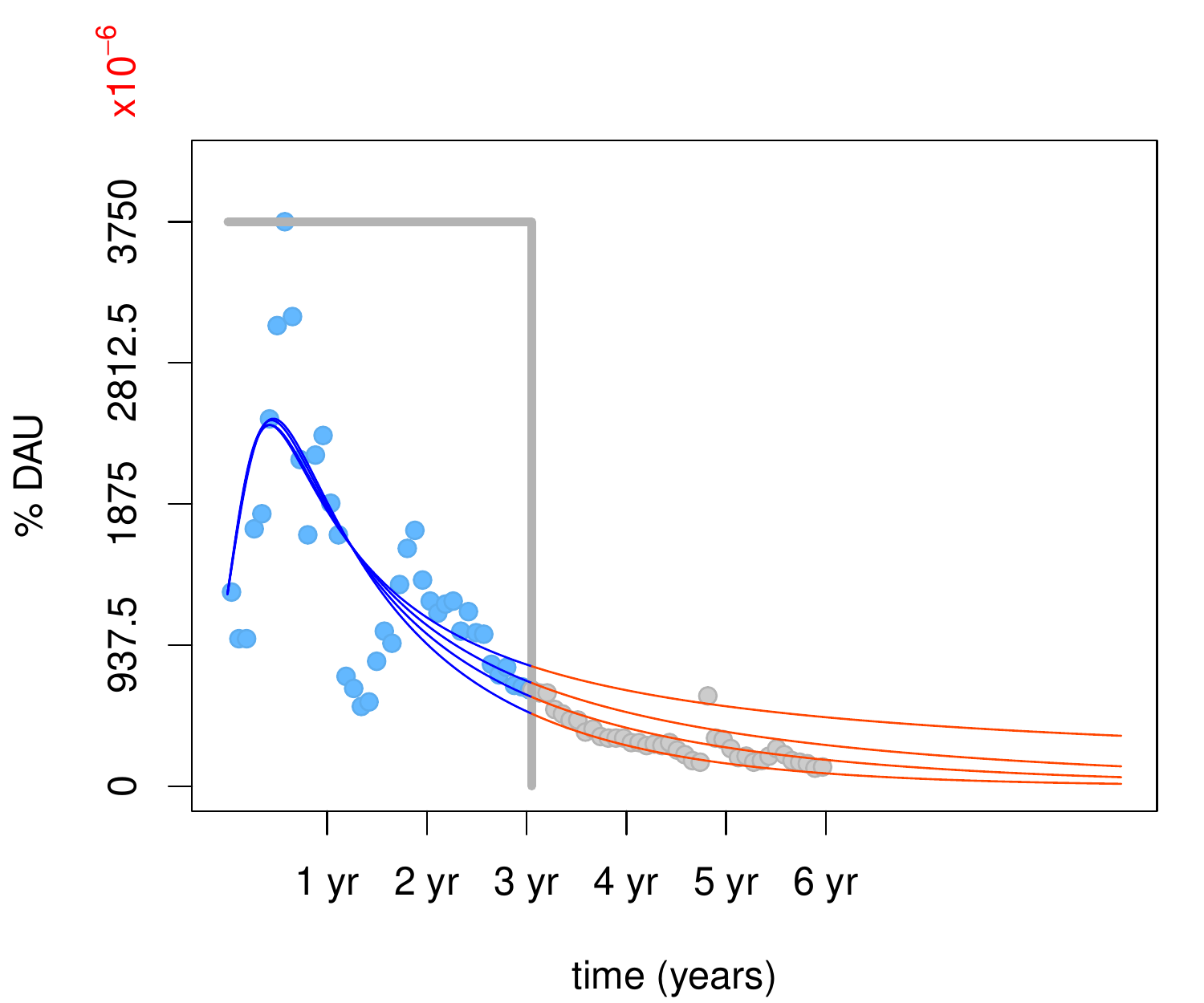}
}
\vspace{-5pt}
\end{center}
\caption{\small %
Incorrect simpler model prediction of flixter.com DAU evolution. Note that unlike our original model, the simpler model presented in {\bf N.B.II} is not able to capture the vanishing DAU trend of flixter. The training data used to generate these graphs is slightly smaller than the one used in our previous experiment.}
\vspace{-10pt}
\end{figure}

In this part of the Appendix we show that the simpler model in {\bf N.B.II} cannot characterize a vanishing DAU time series.
Contrast the prediction for Flixter.com made by the simpler model in Fig.~\subref*{f:simp} against the predictions of our complete model in Fig.~\subref*{f:comp}.
In both cases the training data used to generate these graphs is slightly smaller than the one used in our previous experiment in Fig.~\subref*{f:fli}.
Note that the simpler model shown in Fig.~\subref*{f:simp} incorrectly stabilizes the DAU while the data and our correct model in Fig.~\subref*{f:comp} indicate that the Flixster.com DAU converges towards zero.

\subsection{Dataset description}

The web traffic data used in the evaluation of our model was obtained by the commercial web analytics company Alexa.com, a subsidiary of Amazon.com.
Today, Alexa provides traffic data, global rankings and other information on 30 million websites~\cite{Alexa}.
Alexa ranks sites based primarily on tracking information of users of its toolbar available for all the Internet Explorer, Firefox and Google Chrome  web browsers.
Since 2008 Alexa claims to remove self-selection bias -- bias related to gathering data of a specific audience subgroup that is more likely to install Alexa's toolbar -- by taking into account other data sources ``beyond Alexa Toolbar users''~\cite{Alexa}, but the nature of such data sources and the methodology employed are not disclosed.
Nonetheless, because Alexa's report is detailed and widely used in the industry, we believe that Alexa's {\em unique subscriber} daily traffic reports are a good source of data for our study. The following websites were used:
{\scriptsize
\setlist{nolistsep}
\begin{list}{$\blacktriangleright$}{\leftmargin=1.3em \itemindent=0em}
\setlength{\itemsep}{0pt}%
\setlength{\parskip}{0pt}%
\setlength{\listparindent}{0pt}%
\item {\bf 12seconds.tv:} ``12seconds.tv is a Twitter-like video status service. It gives you 12 seconds to share video moments from your life''~\cite{crunchbase}.
\item {\bf adaptu.com:} Membership-based online mobile wallet.
\item {\bf ashleymadison.com:} ``Ashley Madison is a Married Dating service and social network for those engaged in relationships but looking to have an affair''~\cite{crunchbase}. 
\item {\bf brandstack.com:} ``Brandstack lets designers create their own studios and sell design work to their peers or to directly consumers''. It was sold in December 2011 to DesignCrowd~\cite{crunchbase}.
\item {\bf cafemom.com:} ``CafeMom is a social network site for moms, reaching an audience of more than 20 million users.''~\cite{crunchbase}.
\item {\bf community.babycenter.com:} The Community Baby Center, launched in 2008, is a social network for parents with young children of all ages to share their experiences.
\item {\bf facebook.com:} Online social network website.
\item {\bf flixter.com:} Flixster is a social networking site for movie fans.
\item {\bf formspring.me:} ``Formspring helps people find out more about each other by sharing interesting \& personal responses''~\cite{crunchbase}.
\item {\bf huffingtonpost.com:} The Huffington Post is a leading left-leaning online news aggregator and producer.
\item {\bf linkedin.com:} Is a fast-growing online professional networking website.
\item {\bf marriedsecrets.com:} Married dating service and social network.
\item {\bf meetup.com:} Is a local community social network.
\item {\bf netflix.com:} Membership-based movie rental website.
\item {\bf occupywallst.org:} Is a website for people interested in the Occupy Wall Street movement.
\item {\bf patientslikeme.com:} Patientslikeme is a social networking site that allows people with similar diseases to share their experiences about treatments, doctors, and seek emotional support.
\item {\bf ruelala.com:} Membership-based online retail store.
\item {\bf teapartynation.com \& teapartypatriots.org:} TeaPartyNation.com and TeaPartyPatriots.org are the official website of conservatives American political organizations considered part of the Tea Party movement.
\item {\bf true.com:} True is an online dating service that was founded in 2003~\cite{crunchbase}.
\item {\bf theblaze.com:} The Blaze is a conservative news and opinion website run by conservative pundit Glenn Beck~\cite{TheBlaze}.
\item {\bf webchat.freenode.net:} Is a web-based IRC chat server.
\end{list}
}

\subsection{Model \& the growth of MySpace}
Online social network websites rarely allow access to their subscriber activity data.
To complement the analysis of the preceding sections we use a complementary source of data.
This dataset records the activity of 1.2 million random myspace.com subscribers that joined MySpace from 2006 to 2008 (collected by Ribeiro et al.~\cite{NetSciCom2010}).

~\\
\noindent
{\em A Short History of MySpace.}
MySpace was founded in 2003 and from 2005 until early 2008 MySpace was the most visited social networking website in the world.
In June 2006 MySpace surpassed Google as the most visited website in the United States. 
By late 2007 MySpace first reports of a significant loss of its teenager subscriber base to Facebook appeared~\cite{MySpaceFacebook} and  in April 2008 Facebook usage overtook MySpace~\cite{ListSocial}. 
\\

\begin{figure}[!tb]
\vspace{-.3in} 
\begin{center}
\hspace{-15pt} \includegraphics[width=3.5in,height=1.8in]{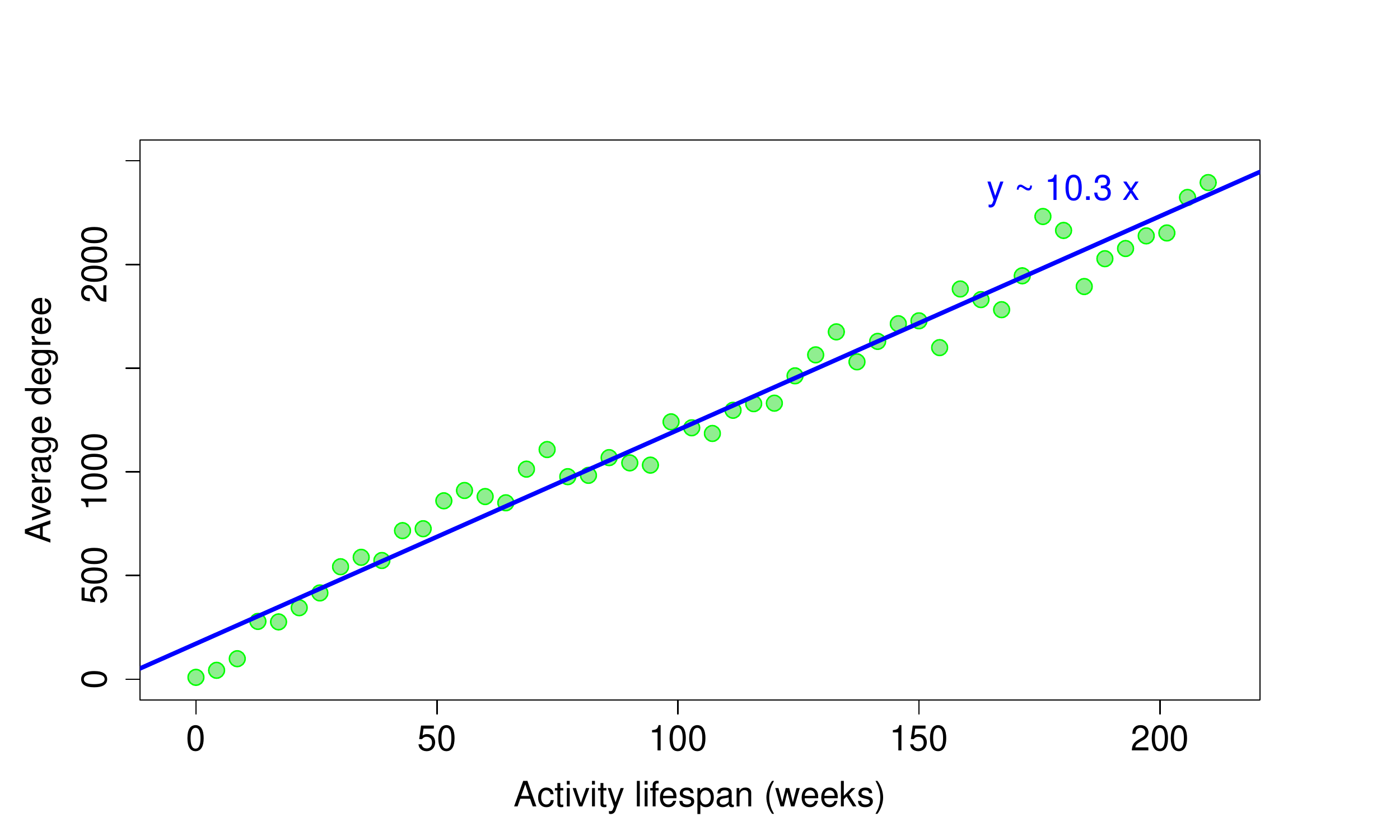}
\end{center}
\vspace{-.2in} 
\caption{\small
MySpace.com average number of friends v.s.\ number of weeks of activity.
Green points show the empirical average. The blue line is the regression showing $y \propto 10.3 x$.\label{fig:myspaceavgac}
}
\end{figure}

\begin{figure}[!bt]
\begin{center}
\includegraphics[width=0.5\textwidth,height=2.0in]{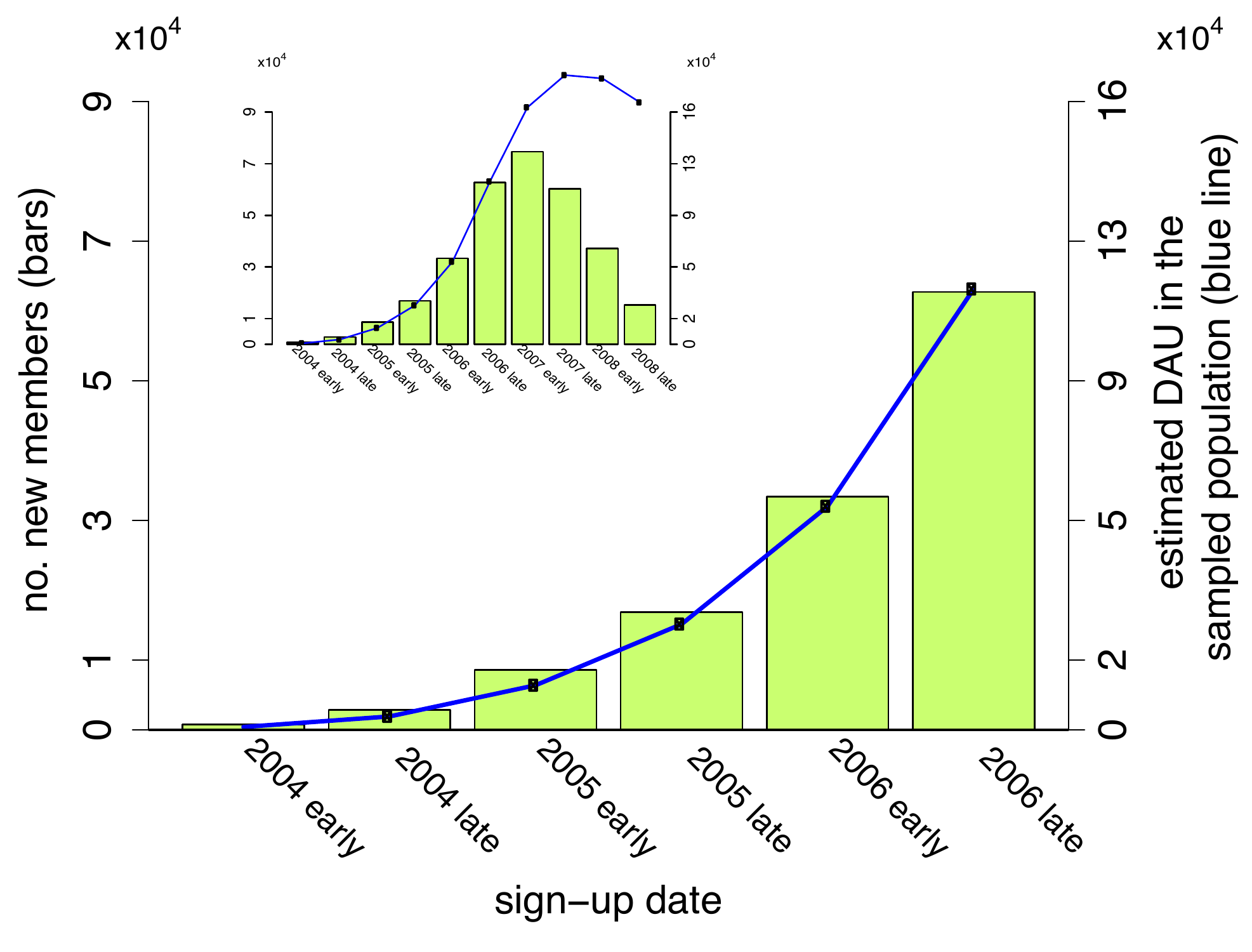}
\end{center}
\vspace{-.6cm}
\caption{%
{\small
{\bf MySpace.com growth $\times$ activity}. 
Linear relationship between number of observed new subscribers (bars) and active subscribers (line) per semester from 2004 to 2006, before MySpace started competing for usage with Facebook. The inset shows the same plot including the years 2007 and 2008 where we observe that while Facebook's competition significantly reduced user activity on MySpace, it has a much milder impact on network growth. \label{fig:join}%
\vspace{-.5cm}
}
}
\end{figure}

We start our analysis with the relationship between network growth and member activity.
The member activity lifespan is defined as the period between the member join date and the member's last login date.
In what follows we make the following reasonable assumption: (a) member lifespan and member activity rate are linearly dependent; and (b) a member active for $x$ semesters has logged in at least once per semester.
In our data we observe that members with an activity lifespan of $x$ weeks have on average $10.3 x$ friends (see Fig.~\ref{fig:myspaceavgac}).
This constant rate of edge acquisition is consistent with measurements in other OSNs~\cite{zhao2012multi,leskovec2008microscopic} and is {\em consistent with our model}.
It is worth noting that Leskovec et al.~\cite{leskovec2008microscopic} measures the rate of edge creation with respect to member's age in the website (the member's age is defined as being the time span between their joining date and the measurement date) rather than member lifetime as we do.
Leskovec et al.\ measurements are less correlated with the DAU than our measurements.

In Fig.~\ref{fig:join} we also see a linear relationship between member activity and the network monthly growth between the first semester of 2004 (early 2004) until the end of the second semester of 2006 (late 2006).
The green bars show the number of new subscribers of each year observed in our random sample while 
the black line shows the number of active subscribers (left vertical axis).
Assuming that the activity rate of a subscriber grows linearly with his or her lifespan, we conclude that between 2004 and 2006 for every active subscriber MySpace acquired a new subscriber.
The above measurement is also {\em consistent with our model}.

\begin{figure}[!ht]
\begin{center}
\hspace{-10pt}
\includegraphics[width=0.45\textwidth,height=1.8in]{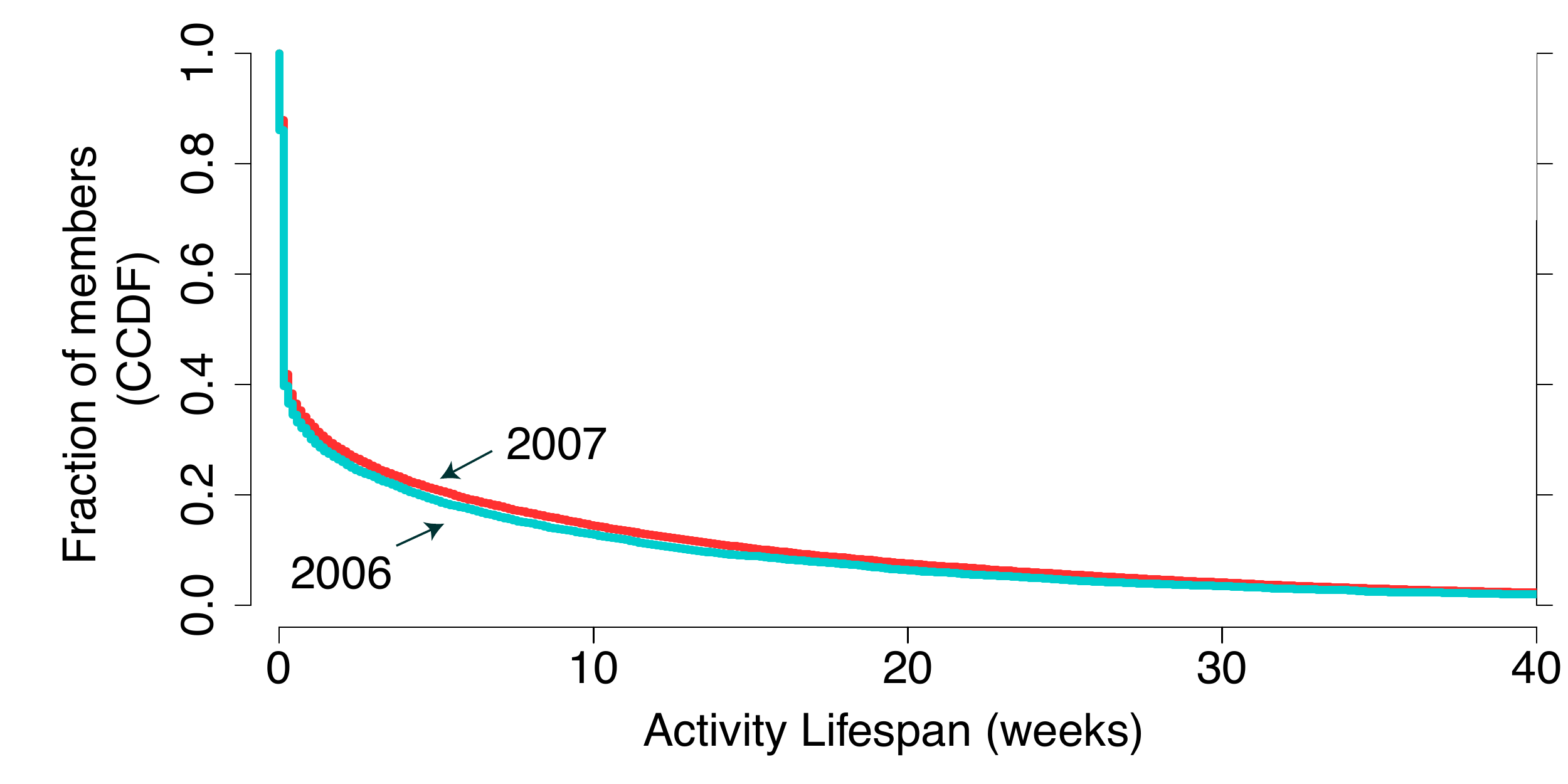}
\end{center}
\vspace{-13pt}
\caption{\small
{\bf  Estimated MySpace subscriber lifespan CCDF}. 
The complimentary cumulative distribution function (CCDF) of the distribution of subscriber activity lifespans.
Estimates obtained by the Kaplan-Meier estimator for subscribers that joined MySpace in 2005, 2006, and 2007 between January--June (early) and July--December (late). We consider subscribers with 60 or more  days of inactivity at the time of measurement to be permanently inactive.\label{fig:mylife}
}
\vspace{-10pt}

\end{figure}

The inset of Fig.~\ref{fig:join} shows the main plot including the years 2007 and 2008. 
Note that the network growth rate (the bars) decreases significantly in 2007 and 2008 when compared to previous years (2004-2006).
Another interesting aspect of the 2007/2008 decline is that -- as seen later in our earlier Technical Report~\cite{techreport} -- Facebook's competition seems to have only significantly affected MySpace by late 2008.
The notion that the Facebook competition may not explain MySpace's falling numbers of new subscribers in 2007  is corroborated by estimating the lifespan distribution between subscribers that joined in 2006 and 2007 using the Kaplan-Meier~\cite{Kaplan:1958cv} estimator (Fig.~\ref{fig:mylife}), which reveals little change in the subscriber lifespan distribution between the years where the number of new members first dropped.

Thus, the unchanging lifespan distribution and the constant activity of subscribers in 2007 both indicate that there is little Facebook effect in the dwindling growth of MySpace. This means that MySpace growth saturated near early 2008 and the fraction of non-members in the population, $U(t)$, started to converge to zero over time and that the member growth is bell shaped in self-sustaining websites, which can be easily verified to be also {\em consistent with our model}.

\balance

\end{document}